\documentclass[11pt]{article}
\input epsf.tex
\newdimen\SaveWidth \SaveWidth=\textwidth
\newdimen\SaveHeight \SaveHeight=\textheight
\advance\SaveWidth by -\textwidth
\advance\SaveHeight by -\textheight
\divide\SaveWidth by 2
\divide\SaveHeight by 2
\advance\hoffset by \SaveWidth
\advance\voffset by \SaveHeight
\def\abs#1{\left| #1\right|}

\def\etal{{\it et al.}}
\def\abs#1{\left| #1\right|}
\def\sgn{\mathop{\rm sgn}}
\def\etmiss{\slashchar{E}_T}\def\TeV{{\rm TeV}}

\def\pb{{\rm pb}}

\def\fbi{{\rm fb}^{-1}}

\def\fbi{{\rm fb}^{-1}}
\def\Meff{M_{\rm eff}}

\def\lsp{{\tilde\chi_1^0}}

\def\GeV{{\rm GeV}}
\def\mhalf{m_{1/2}}
\def\tchi{\tilde\chi}

\def\tq{\tilde q}

\let\badcite=\cite
\def\cite{~\badcite}

\def\slashchar#1{\setbox0=\hbox{$#1$}           
   \dimen0=\wd0                                 
   \setbox1=\hbox{/} \dimen1=\wd1               
   \ifdim\dimen0>\dimen1                        
      \rlap{\hbox to \dimen0{\hfil/\hfil}}      
      #1                                        
   \else                                        
      \rlap{\hbox to \dimen1{\hfil$#1$\hfil}}   
      /                                         
   \fi}                                         %

\catcode`@=11
\newdimen\vbigd@men                             

\def\vbig#1#2{{\vbigd@men=#2\divide\vbigd@men by 2%
   \hbox{$\left#1\vbox to \vbigd@men{}\right.\n@space$}}}
\def\dofigs#1#2#3{\centerline{\epsfxsize=#1\epsfbox{#2}%
   \hfil\epsfxsize=#1\epsfbox{#3}}}
\catcode`@=12

\catcode`@=11
\def\citenum#1{\csname b@#1\endcsname}
\catcode`@=12

\def\dofig#1#2{\centerline{\epsfxsize=#1\epsfbox{#2}}}
\begin{document}
\begin{titlepage}
\rightline{ATL-COM-PHYS-2001-033}
\rightline{\today}
\begin{center}{\Large\bf Impact of Energy and Luminosity
    upgrades at LHC on the Physics program of ATLAS}
\end{center}
\bigskip
\begin{center}
\bf  G.~Azuelos$^a$,   D.~Benchekroun$^b$,
O.~\c{C}ak\i r$^c$, E. Elfgren$^a$,
F.~Gianotti$^d$,  J.-B.~Hansen$^d$, , I.~Hinchliffe$^e$, M.~Hohlfeld$^f$, K.~Jakobs$^f$,
C.~Leroy$^a$, 
R.~Mehdiyev$^{a,k}$, F.E. Paige$^g$, 
 G.~Polesello$^h$, H.~Stenzel$^i$, 
  S.~Tapprogge$^j$, Z.~Usubov$^{k,l}$,   L.~Vacavant$^e$
\end{center}
\centerline{$^a${\it U. of Montreal, Montreal, Canada}}
\centerline{$^b${\it  U. Hassan II Casablanca-Maarif, Casablanca. Morocco}}
\centerline{$^c${\it U. of Ankara, Turkey}}
\centerline{$^d${\it CERN, Geneva, Switzerland}}
\centerline{$^e${\it Lawrence Berkeley National Laboratory, Berkeley,
    CA}}
\centerline{$^f${\it Institut fur Physik, Johannes Gutenberg-Univeritat
    Mainz, Germany}}
\centerline{$^g${\it Brookhaven National Laboratory, Upton, NY}}
\centerline{$^h${\it INFN, Sezione di Pavia,  Pavia, Italy}}
\centerline{$^i${\it Max-Planck-Institut fur  Physik, Munich, Germany}}
\centerline{$^j${\it HIP, Helsinki, Finland}} 
\centerline{$^k${\it Institute of Physics, Academy of Sciences of
    Azerbaijan, Baku, Azerbaijan}}
\centerline{$^l${\it Joint Institute for Nuclear Research, Dubna,
    Russia}}

 \bigskip

\begin{abstract}
The impact on the physics capabilities of the ATLAS detector of
possible LHC upgrades is discussed. As a benchmark, an 
increase in the luminosity by a factor of ten is
considered. For comparison, a  doubling of the LHC energy is also explored.
Both upgrades significantly enhance the physics
capabilities of ATLAS. As measured in terms of the mass reach for new
particles, the energy upgrade is more powerful. However, in cases where the
effect of an upgrade is to increase the precision of measurements as a
result of the larger data samples, the luminosity upgrade can be 
at least as powerful.  The pile-up
of minimum bias events at higher luminosity could limit the physics
performance of ATLAS in areas where tagging of forward jets is needed.
\end{abstract}
\end{titlepage}

\section{Introduction}

It is useful to consider the possible impact on physics of upgrades to
the LHC which would permit operation at a higher luminosity. For
comparison purposes, the effect of an energy increase will also be considered.
Four  basic scenarios  for high energy/luminosity $pp$
colliders are 
used in this note.
The higher luminosity operation would correspond to a bunch spacing
of 12.5 ns. 
\begin{itemize}
\item A The approved LHC, {\it i.e.} $\sqrt{s}=14$ TeV and luminosity of $10^{34}$
  cm$^{-2}$ sec$^{-1}$
\item B $\sqrt{s}=14$ TeV and luminosity of $10^{35}$
  cm$^{-2}$ sec$^{-1}$
\item C $\sqrt{s}=28$ TeV and luminosity of $10^{34}$
  cm$^{-2}$ sec$^{-1}$
\item D $\sqrt{s}=28$ TeV and luminosity of $10^{35}$
  cm$^{-2}$ sec$^{-1}$
\end{itemize}

Note that, while scenario B represents a realistic upgrade scenario for the
 LHC, scenarios C and D are much more speculative. 
The impact of the upgrades on detector performance has not been
assessed in detail but   a few general remarks are appropriate. The
baseline performance is documented in \cite{TDR}. The
higher luminosity has a  severe impact. In the case of the ATLAS Inner
Detector, the Transition Radiation Detector  
would need to be replaced due to the high
occupancy. Radiation damage would imply replacement or removal of the
pixels and some  of the silicon layers. Removal would imply loss of
tracking, electron identification
and b-tagging capability. Calorimeter performance is not expected to
be impacted significantly in the central region; some of the
Tile Calorimeters' scintillating
 fibers may need
to be replaced. The forward calorimeter (FCAL) could suffer from  
space-charge effects. 
More shielding will  be required for the forward muon system which
will reduce the 
fiducial region. 
The reduced bunch spacing at very high luminosity could have a severe impact
on the front-end
electronics, trigger and DAQ components. For example, the front
end pipeline buffers would be incompatible with the latency needed for
the level-1 trigger. The level-1 muon trigger could also be
compromised by increased occupancy in the Resistive Plate  and Thin
Gap Chambers.

The ATLAS B-physics programme \cite{TDR} does not benefit
significantly from the energy or
luminosity upgrades. The programme is accomplished primarily at low
luminosity and, as the total B cross section increases only slightly at the
higher energy, the energy upgrade will have little impact. 
Similarly the top production rate is very large in the baseline LHC 
and, possibly 
apart from rare decay modes, the physics will be fully exploited at
design energy/luminosity. 

Physics processes involving the production of high-mass systems have
cross-sections that rise rapidly with energy. While the luminosity
upgrade 
has the capability to significantly
 enhance the LHC capability, one  should expect
therefore that, if the impact of upgrades is assessed in terms of mass 
reach for new particles, the energy upgrade will be more powerful.
For cases involving somewhat smaller masses, the luminosity upgrade
could be more powerful {\it provided that the detector performance can
  be maintained.} In subsequent sections, various high $p_T$
physics cases are studied ranging from QCD to
extra-dimensions. Finally some conclusions are drawn.

\section{QCD studies and searches for compositeness}
  
\begin{figure}
\dofig{3in}{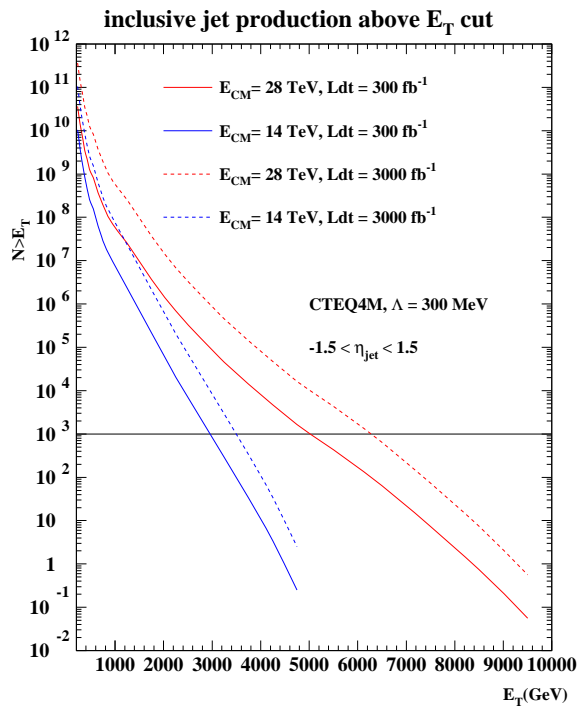}
\caption{
  Expected number of events with a jet of transverse energy larger
  than $E_T$ for the four scenarios studied.
\label{jets}}
\end{figure}

For QCD studies at an upgraded LHC, a comparison of the reach in jet
  transverse momenta was carried out using a NLO calculation of the
  inclusive jet cross section \cite{JETRAD}. The expected number of
  events with a jet above a transverse energy  threshold is shown in 
Figure~\ref{jets}.
 Requiring that there be 1000 events above a threshold,
  the accessible $E_T$ is shown in Table~\ref{jettab}.
 This comparison clearly shows the advantage of increasing the
  center-of-mass energy to 28 TeV over a 10-fold increase in
  luminosity. In the former case, the reach extends by about 
  2 TeV, larger by a factor of four than  in the latter 
  case. As the measurements involve the calorimeters only and 
  only  jets in the TeV range are of interest,
  the detector performance is not expected to be
degraded significantly at the highest luminosity.

\begin{figure}
\dofig{3in}{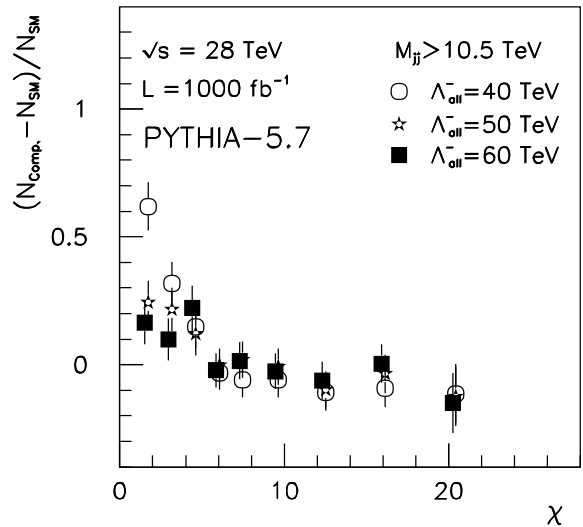}
\caption{Expected deviation from the Standard Model
  predictions for the angular distribution of dijet pairs at 28 TeV
  for various values of the compositeness scale $\Lambda$. 
Dijet pairs
  are required to have invariant mass greater than 10.5 TeV. 
\label{comp1}}
\end{figure}

Signals for quark compositeness should reveal themselves in deviations of the 
high energy part of the jet cross-section from the QCD
expectation. The angular distribution of dijet pairs of large
invariant mass  provides an
independent diagnostic   and is less sensitive to possible non-linearities
in the calorimeter response (see Section 21.5 of Ref~\cite{TDR} for more
details on the methodology). It was therefore used in this study.   Figure~\ref{comp1} shows the distribution
for dijet events in the variable 
$\chi=(1+\abs{\cos\theta})/(1-\abs{\cos\theta})$,
where $\theta$ is the angle between a jet and the beam in the
center-of-mass frame of the dijet system.  The invariant mass of the
dijet system is required to be more than 10.5 TeV. Shown is the deviation from
the expected form in the Standard Model for various values of the
compositeness scale $\Lambda$ at $\sqrt{s}=28 $ TeV. 
 The effect of compositeness shows up as an increase in
the event rate at small values of $\chi$. 
The compositeness scales that can be probed  in this
manner are   shown in 
Table~\ref{comptab}. As this comparison of the scenarios is based on
the production of the very highest energy particles, the energy
upgrade more sensitive.
\begin{table}
\begin{center}
  \begin{tabular}{|c|c|c|c|c|}\hline
      scenario         &    14 TeV 300 fb$^{-1}$&  14 TeV 3000 fb$^{-1}$
&28 TeV 300 fb$^{-1}$ &28 TeV 3000 fb$^{-1}$ \cr\hline
      max $E_T$ (TeV)  &     3.0  &     3.5 &      5.0   &    6.3\cr 
\hline\end{tabular}\end{center}
\caption{The maximum value of jet transverse momentum that is
  accessible in various LHC energy/luminosity scenarios. The criterion
  is that there be 1000 events with $p_T$ greater than the value shown.
}\label{jettab}
\end{table}

\begin{table}
\begin{center}
  \begin{tabular}{|c|c|c|c|c|}\hline
      scenario         &    14 TeV 300 fb$^{-1}$&  14 TeV 3000 fb$^{-1}$
&28 TeV 300 fb$^{-1}$ &28 TeV 3000 fb$^{-1}$ \cr\hline
      $\Lambda$(TeV)  &     40  &     60 &     60   &    85\cr 
\hline
  \end{tabular}\end{center}\caption{The 95\% confidence level limits
  that can be obtained on the compositeness scale
  $\Lambda$}\label{comptab} using the dijet angular distributions.
\end{table}

\section{Higgs Boson studies}

\begin{figure}
\dofig{3in}{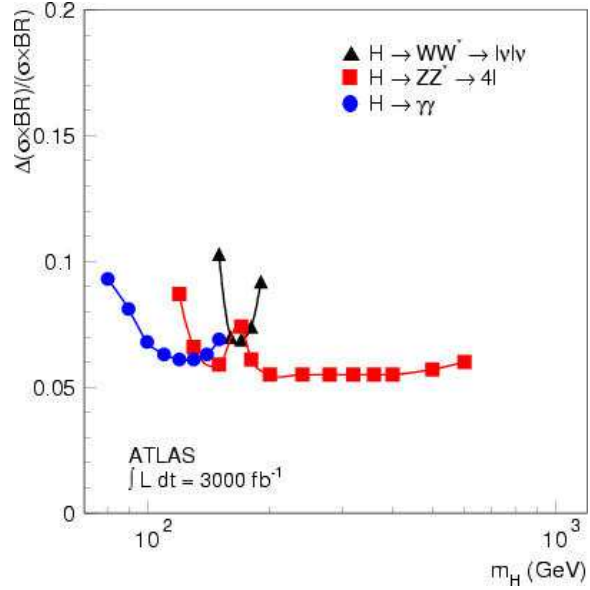}
\caption{Expected precision on the  measurement of Higgs boson
production rates for  at 14 TeV and very high
  luminosity.
\label{karl1}}
\end{figure}

\begin{figure}
\dofig{3in}{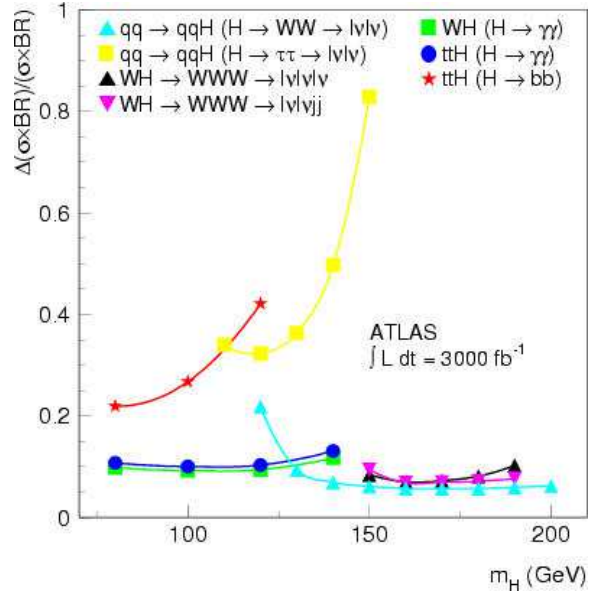}
\caption{
Expected precision on the measurements of the production rates  of
  the Higgs boson in various channels at 14 TeV and very high
  luminosity.
\label{karl2}}
\end{figure}

The Higgs physics discovery program will largely be 
completed before any upgrade. Following the discovery, 
the measurement of the mass will have
reached the systematic limit, whereas
measurements of couplings will be limited by the available statistics
and could benefit from an upgrade.
The following final states  are expected to be observable depending upon
the mass: $\gamma\gamma$, $\tau\tau$,
$ZZ\to \ell\ell\ell\ell$, $WW\to\ell\nu\ell\nu $ and $b\overline{b}$. 
The relevant
production processes are $qq\to qqH$ (the vector boson 
fusion process) which 
depends upon the $HWW$ and $HZZ$ couplings, $gg\to H$ via a loop
involving, in particular,  top quarks, $q\overline{q}\to HW$, 
and $gg\to t\overline{t}H$
which is sensitive to the $ttH$ coupling. Studies of the expected
precision for the benchmark LHC  \cite{Hohlfeld}
 have been extended for the upgrade scenarios.  The  expected precision on
the production rates is shown in Figures~\ref{karl1} and
\ref{karl2}. In these plots it is assumed that there is an error of
5\% on the measurement of the absolute luminosity. In some cases this
becomes the limiting factor.  Figure~\ref{karl1} shows the measurements for
  final states containing gauge bosons  produced by 
the gluon-gluon fusion 
 process. Figure~\ref{karl2} shows that for the other processes.
 In the mass range below 200 GeV which is favored
 by fits to electroweak data  \cite{lepfits}, there are a large number of possible
 measurements. The factor of ten increase in luminosity over the
 standard LHC scenario improves the precision by approximately a
 factor of two and brings many of these measurements 
 to the point where they are
 limited by the uncertainty on  the measurement of the luminosity and
 other systematic effects. In particular, we have assumed that the
 measurement of 
 $tau\tau$ final state does not improve at very high luminosity due to
 the difficulties of reconstructing these states. Also the $WW$
 final state is compromised by the background as it has no mass peak
 and the uncertainty in the shape of this background contributes to the
 uncertainty.

\begin{figure}
\dofig{3in}{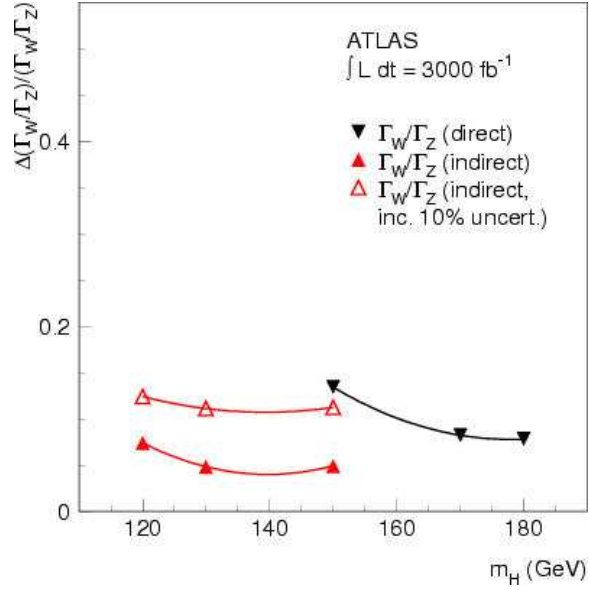}
\caption{
Expected fractional error on the measured ratio of Higgs partial widths to $WW$ and
$ZZ$ as a function of the Higgs mass. The results labeled direct are
obtained with no additional theoretical assumptions. Those
labeled indirect have additional assumptions (see text). Also shown
is the effect of a 10\% uncertainty in the theoretical assumptions.
\label{karl3}}
\end{figure}

 Once these cross-section measurements are made, they can be
 used to extract information on the couplings of the Higgs boson. In
 the case where measurements of different final states are made from
 the {\it same} Higgs production mechanism, a direct measurement of
 the ratios of Higgs couplings is possible.  Figure~\ref{karl3} shows
 the expected precision on the ratio of the Higgs partial widths to the $WW$
 and $ZZ$ final states. For masses greater than approximately 160 GeV
 measurements  of  $H\to ZZ\to 4\ell$ and  $H\to WW\to \ell\nu\ell\nu$
 rates from the same production process enables a direct (model independent)
 measurement to be made. At smaller masses the process  $H\to WW\to
 \ell\nu\ell\nu$ has too low a rate but an indirect  measurement can be obtained
 by combining the measured rates for  $H\to ZZ\to 4\ell$ and
 $H\to\gamma\gamma$ and making a theoretical assumption.
 The coupling  $H\to\gamma\gamma$
 is dominated by a loop graph with an intermediate $W$ and hence the
 rate $H\to\gamma\gamma$ can be related to the $HWW$ coupling.

\begin{figure}
\dofig{3in}{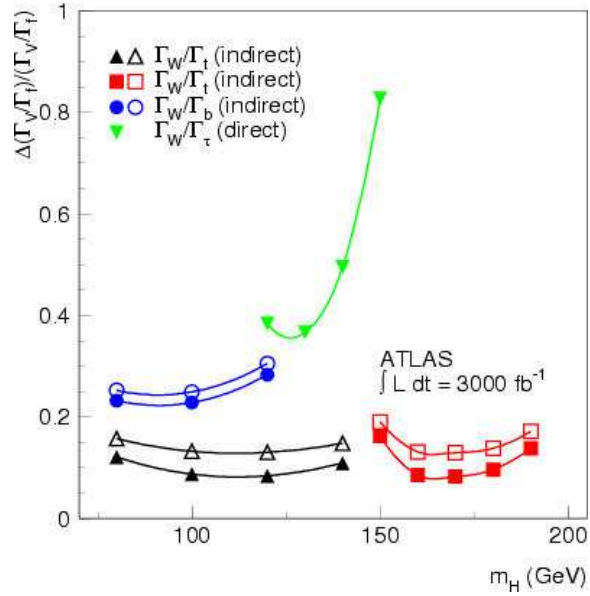}
\caption{Expected fractional error on the measured ratio of Higgs
  partial  widths to
  several final states as a function of the Higgs mass. The widths
  shown are $\tau\tau$, $WW$  and $b\overline{b}$. The results 
labeled direct are
obtained with no additional theoretical assumptions. The top
quark coupling strength $\Gamma_t$, is inferred from other measurements
(see text). The open symbols contain an additional 10\% theoretical uncertainty.
\label{karl4}}
\end{figure}

 Measurements of the fermion couplings shown in Figure~\ref{karl4} are
 more difficult. Events with  
  $H\to \tau\tau$ and
 $H\to WW\to\ell\nu\ell\nu$ and  tagged forward jets
 arise from the same production process ($qq\to qqH$) and hence
 can be combined to directly measure $\Gamma_W/\Gamma_{\tau}$ as shown in
 Figure~\ref{karl4}. The coupling $Ht\overline{t}$ can be probed by
 comparing the $WH\to \ell\nu\gamma\gamma$ and $H\to\gamma\gamma$
 rates. The latter production rate is determined by the coupling of
 Higgs to gluon pairs and the dominant contribution to this coupling
 is from a top quark loop which probes the $Ht\overline{t}$ coupling
 strength. An indirect measurement of the ratio of couplings $HWW$ and
 $Ht\overline{t}$ can therefore be made and is shown on Figure~\ref{karl4}
Similarly the final states $t\overline{t}H(\to\gamma\gamma)$ and
 $t\overline{t}H(\to b\overline{b})$ can be combined to give the ratio
 of widths to $b\overline{b}$ and $WW$ shown
 as an indirect  measurement
 in Figure~\ref{karl4}.
  Again the increased luminosity improves the precision of the measurements by
  approximately a factor of two.

If the energy of the LHC were raised to 28 TeV, then for a Higgs boson
of mass $\sim 150$ GeV ($\sim 1$ TeV), 
the Higgs production cross-section would rise by a factor of 2
(8). Therefore, in the case of Higgs masses at the lower end of the
search range, the luminosity upgrade is much more powerful. However,
it is important to emphasize that the full detector functionality and 
performance must be
maintained at the upgraded luminosity if this advantage is to be exploited.

In the case of SUSY Higgs bosons there are  regions of the MSSM  parameter 
space where only the lightest Higgs  state ($h$) will have been 
seen at LHC.
 In this case the sensitivity to the heavier Higgs states via for the
 decays $A \to \tau\tau$ and $A\to \mu\mu$ will be increased at an
 upgraded LHC. 
In the case of $m_A=500$  GeV the energy upgrade increases the H/A
cross-section by approximately a factor of five thereby increasing the
discovery potential for heavy Higgs bosons.

\section{Supersymmetry studies}

Perhaps the strongest argument for weak scale SUSY is that it allows
 light Higgs bosons without unnatural fine tuning in the
presence of much heavier scales such as the Planck mass. While the
concept of fine tuning is not precise, there is a general consensus that
this requires SUSY particles to be lighter than about $1\,\TeV$; see,
{\it e.g.}, \cite{Anderson:1995r}. Masses in this range are also
consistent with grand unification, cold dark matter, an observable SUSY
contribution to $g_\mu-2$~\cite{Brown:2001mg,Czarnecki:2001pv},
and a variety of other
constraints~\cite{Ellis:2001yu,Everett:2001tq,Baer:2001kn,Martin:2001st,Feng:2001tr}.

If $R$ parity is conserved, all SUSY particles decay to the lightest
SUSY particle (LSP) $\lsp$, which escapes the detector. Thus, the
production of gluinos or squarks provides signatures with multiple jets,
possibly leptons, and missing transverse energy $\etmiss$. In typical
SUSY breaking scenarios such as minimal SUGRA \cite{sugra}, the LHC should discover
evidence for gluinos or squarks up to $1\,\TeV$ in the $\etmiss + {\rm
jets}$ channel with an integrated luminosity of $1\,\fbi$ or less; the
ultimate discovery reach for $300\,\fbi$ extends beyond $2.5\,\TeV$.

\begin{figure}
\dofig{3in}{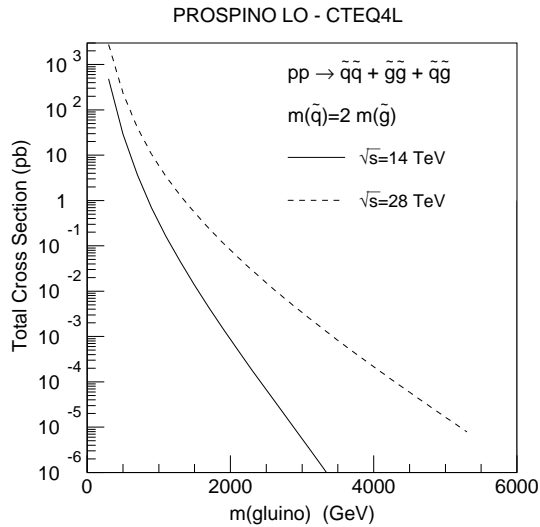}
\caption{
  The total SUSY production cross section as a function of the gluino mass for
  the case where $m_{\tilde{q}}=2.0m_{\tilde{g}}$.
\label{gg}}
\end{figure}

\begin{figure}
\dofig{3in}{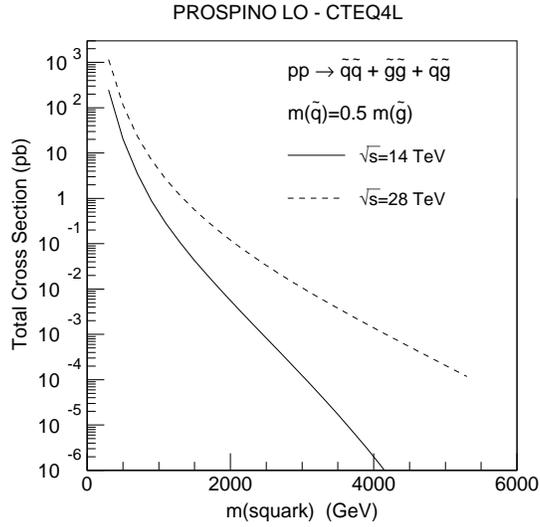}
\caption{
  The total SUSY production cross section as a function of the squark  mass for
  the case where $m_{\tilde{q}}=0.5m_{\tilde{g}}$.
\label{qq}
}
\end{figure}

\begin{figure}
\dofig{3in}{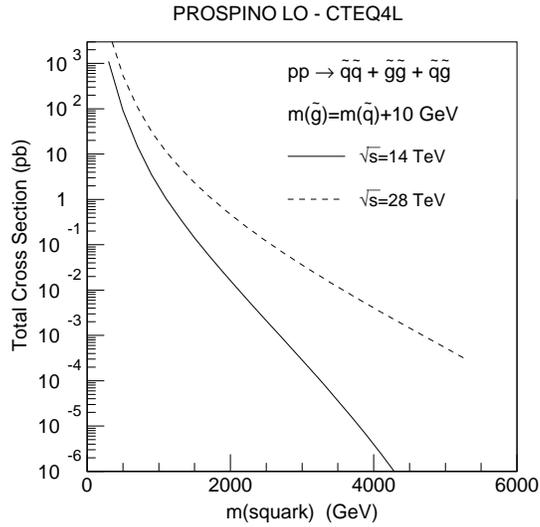}
\caption{
  The total SUSY production cross section as a function of the squark  mass for
  the case where  $m_{\tilde{g}}=m_{\tilde{q}}+10$ GeV.
\label{qg}}
\end{figure}

The increase in mass reach for squarks and gluinos at higher LHC
energy 
is  evaluated  approximately by comparing the squark and gluino 
production cross-sections at 14 and 28 TeV.
The primary Standard Model backgrounds,
$t\overline{t}$  production and W/Z+jets increase due to the 
larger center of mass energy. However,  
 this effect will be more than compensated
by the harder cuts on the transverse energy flow and missing $E_T$ 
in the events which can be applied  compared to the analysis 
performed for a 14 TeV collider.
The cross-section  was evaluated  at leading order 
using the program PROSPINO \cite{prospino} with  CTEQ4L\cite{cteq} structure 
functions. Three scenarios were considered:
\mbox{$m_{\tilde q} = 2 m_{\tilde g}$}, 
\mbox{$m_{\tilde g} = 2 m_{\tilde q}$},
\mbox{$m_{\tilde g} \sim  m_{\tilde q}$}. 
The total cross section 
 for the production of squarks and gluinos
is given in the Figures~\ref{gg}, \ref{qq} and \ref{qg}
 as a function of the mass of the 
lighter sparticle be it squark or gluino in these scenarios.


 An
 approximation  of the reach at 28 TeV is obtained 
by taking
the masses for which the signal cross-section is equal to 
the one for the maximum reachable masses at 14~TeV\cite{TDR}, {\it i.e.} 
around $10^{-2}$ pb for 100 fb$^{-1}$. 
The approximate SUSY mass reach  for an integrated 
luminosity of 100 fb$^{-1}$ at 28 TeV is thus:    
3.9~TeV for $m_{\tilde q} \sim  m_{\tilde g}$,
2.7~TeV for \mbox{$m_{\tilde q} = 2 m_{\tilde g}$}, 
and 2.9 TeV for \mbox{$m_{\tilde g} \sim  2 m_{\tilde q}$},
which corresponds to an approximate doubling of the accessible mass
range.
An increase in the luminosity is somewhat less powerful as can be seen
from Table~\ref{susytab}, but it still provides a mass reach 20\% higher than
the standard LHC scenario reaching 2.7 TeV for the case $m_{\tilde q} \sim  m_{\tilde g}$. 

\begin{table}
\begin{center}
\begin{tabular}{|c|c|c|c|c|}
\hline
scenario   & 14 TeV  & 14 TeV & 28 TeV & 28 TeV
\cr
 & 100 fb$^{-1}$ & 1000 fb$^{-1}$
 & 100 fb$^{-1}$ & 1000 fb$^{-1}$\cr
\hline
$m_{\tilde q} \sim  m_{\tilde g}$& 2.1 &2.7  &3.9 & 4.4\cr
$m_{\tilde q} = 2 m_{\tilde g}$ &2.0&2.3&2.7 &4.2 \cr
$m_{\tilde g} \sim  2m_{\tilde q}$& 1.9 &2.2  &$2.9$& 4.1\cr
\hline\end{tabular}
\end{center}
\caption{Comparison of the reach in squark masses in TeV in the
  different scenarios. In each case  a nominal one year of running
  is 
assumed.\label{susytab}}

\end{table}

Given the naturalness arguments made above it is
almost guaranteed that SUSY will be discovered  at the baseline LHC 
if it is relevant to the
problem of electroweak symmetry breaking. 
Thus, the  role of an LHC upgrade in energy, luminosity, or both would
be to
complete the spectrum and 
make additional precise measurements. What measurements might be of
interest many years after the first discovery of SUSY is of course model
dependent and difficult to predict. It seems likely, however, that these
measurements will require the full capability of ATLAS  including good identification and measurements of $e$'s,
$\mu$'s, $\tau$'s, jets, $b$-jets, and $\etmiss$.

While gauginos are likely to be lighter than $1\,\TeV$, sfermions might
be heavier. In the ``inverted hierarchy'' scenario~\cite{Bagger:1999ty},
all scalars are heavy at the GUT scale. With appropriate GUT boundary
conditions the renormalization group equations drive the third
generation sfermions below $1\,\TeV$ but leave the rest heavy, perhaps
several TeV. In the ``focus point'' scenario~\cite{Feng:2000mn}, even
the stops ($\tilde t_{1,2}$) can have multi-TeV masses. In scenarios like these,
gluinos and their decay products would still be discovered quickly at
the LHC, but some or all of the squarks and sleptons would be
inaccessible.  A search for very massive squarks might impose fewer
constraints on the detector performance than precision measurements at
very high luminosity, but no simulation has been
done. Models with very massive sfermions often lead to many competing
gluino decay modes, each with a small branching ratio. Precision
measurements of gluino processes at the LHC could be rate limited 
 and might also benefit by an upgrade.

\begin{figure}[t]
\dofig{3in}{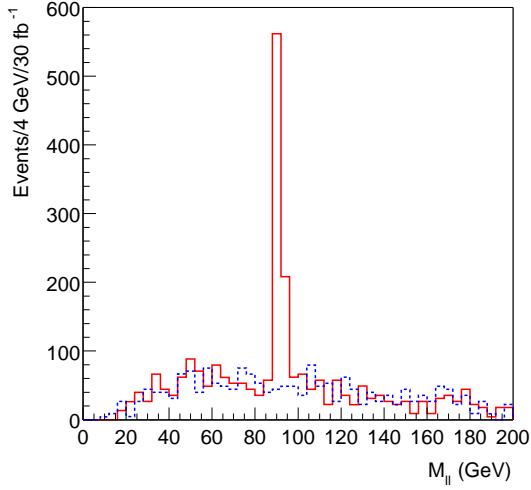}
\caption{$\ell^+\ell^-$ mass distribution at SUGRA Point 2 at $28\,\TeV$.
\label{c228mll}}
\end{figure}

\begin{figure}[t]
\dofig{3in}{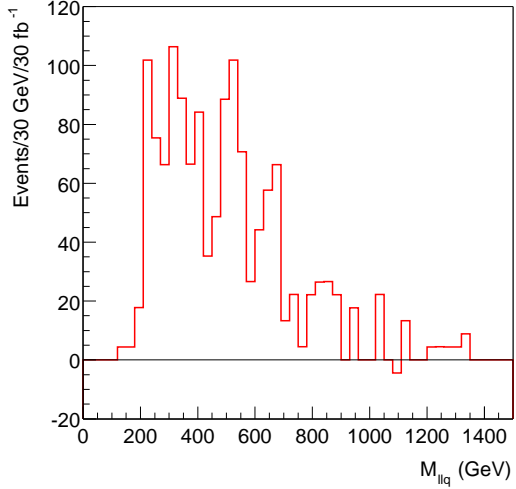}
\caption{$\ell^+\ell^- j$ mass distribution formed from $\ell^+\ell^-$
in the $Z$ mass peak plus one of the two hardest jets at SUGRA Point 2
at $28\,\TeV$. \label{c228mllq}}
\end{figure}

\begin{figure}[t]
\dofigs{3in}{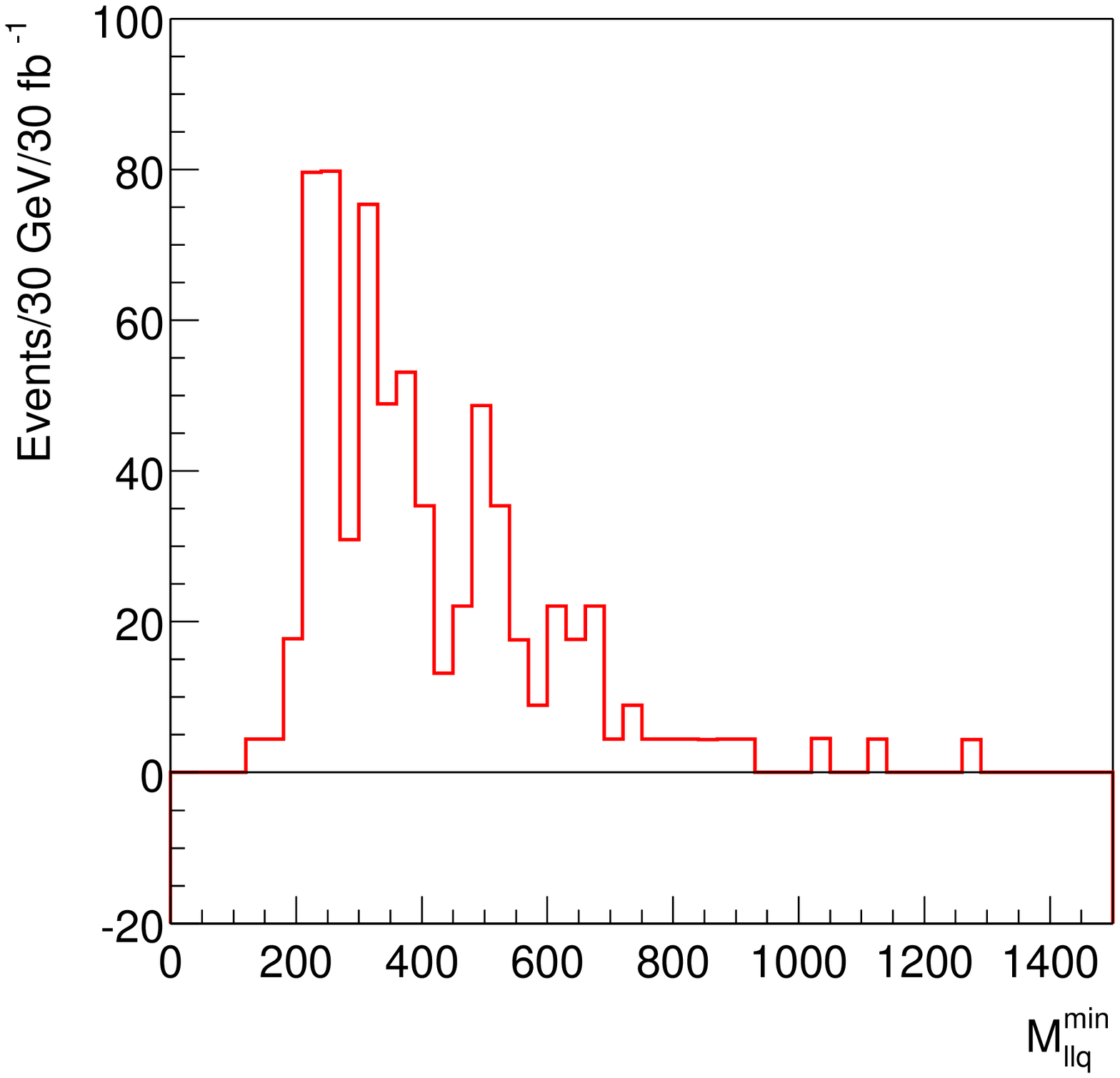}{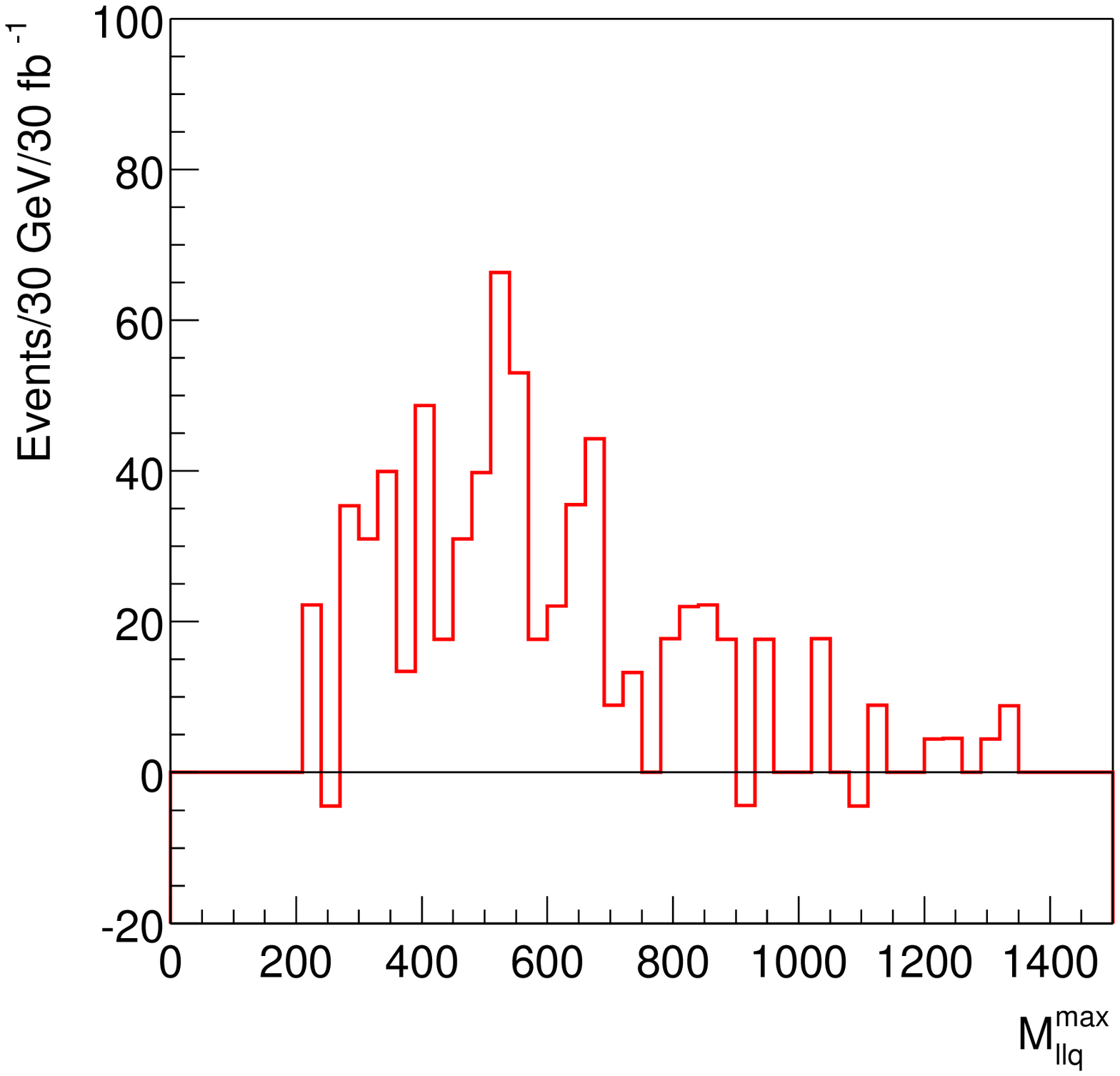}
\caption{Minimum (left) and maximum (right) masses selected as in
Fig.~\protect\ref{c228mllq}. \label{c228mllqmn}}
\end{figure}

SUGRA Point~2 from \cite{TDR} was chosen
as an example of the possible gains for SUSY studies from an energy
or luminosity upgrade of the LHC. This point has $m_0=\mhalf=400\,\GeV$, $A_0=0$,
$\tan\beta=10$, and $\sgn\mu=+$, giving gluino and squark masses of
about $1\,\TeV$. While discovery of a signal in the $\etmiss + {\rm
jets}$ channel is straightforward, the precision measurements discussed
in Ref.~\cite{TDR} were relatively difficult. The channel
$\tchi_2^0 \to \lsp Z$, $Z \to \ell^+\ell^-$ was not used 
 because of the small rate as can be seen in  Fig.~20-25 of
 Ref~\cite{TDR} which  shows
an inclusive sample of 70 events for an integrated luminosity of 
$30\,\fbi$. A luminosity upgrade
would clearly result in an event sample large enough to do a detailed
study provided the detector performance can be maintained.
At $28\,\TeV$ the total
SUSY production cross section is increased to $37\,\pb$ from
$3.2\,\pb$ at 14 TeV showing that in this example the luminosity and
energy upgrades are almost the same  in their effect on the production
rates. In an attempt to exploit the larger event
rates,
 events were selected with
cuts similar to those used in \cite{TDR} using ISAJET 7.51 \cite{ISAJET}:
\begin{itemize}
\item   At least 4 jets with $p_T>100,50,50,50\,\GeV$;
\item   $\Meff>1000\,\GeV$;
\item   $\etmiss>0.2\Meff$;
\item   $S_T>0.2$.
\end{itemize}
where $$ \Meff = p_{T,1} + p_{T,2} + p_{T,3} + p_{T,4} + \etmiss\,.  
$$ and  $S_T>$ is the transverse sphericity.
Standard Model backgrounds were not generated, but they should be small
after these cuts are made, as shown in \cite{TDR}. Leptons
($e$,$\mu$) were required to have $p_T>10\,\GeV$ and to be isolated with
$E_T<10\,\GeV$ in  a cone of $R=0.4$ around the lepton direction. The dilepton mass distribution with these
cuts is shown in Fig.~\ref{c228mll}. The $Z\to \ell^+\ell^-$ peak now
has more than 500 events for an integrated luminosity of $30 $ $ \fbi$.
Thus either the energy or luminosity upgrade produces an event sample
large  enough to perform further analyses.

For example, the main source of $\tchi_2^0$ at Point 2 is $\tq_L \to \tchi_2^0 q$.
The two hardest jets in SUSY events are expected to come from $\tq$
decays. The $\ell^+\ell^- q$ mass distribution formed by combining the
two leptons in the $Z$ mass peak with each of the two hardest jets is
shown in Fig.~\ref{c228mllq}. The smaller of these masses should be less
than the kinematic upper limit for $\tq_L \to \tchi_2^0 q \to \lsp Z
q$, while the larger should be greater than the kinematic lower limit.
These distributions are shown in Fig.~ \ref{c228mllqmn}. In either
upgrade scenario 
the statistics are sufficient to make 
measurements that significantly constrain the squark and  $\tchi_2^0$
masses in a manner that would not be possible with the baseline  LHC.


\section{Jet tagging/vetoing at higher luminosity}

\label{sec:tag}
The presence of jets at large rapidity can be used to enhance the
signal-to-background ratio in certain processes such as those
involving $WW$ fusion  where the $W$'s are emitted from
quarks (see, for example, 19.2.10.1 of Ref \cite{TDR}). Similarly the
absence
of jets at central rapidity  can be used to enhance processes
involving the production of particles with only electroweak charges
relative to the production of strongly coupled particles (see, for
example, section 19.2.10.2 of Ref \cite{TDR}). As the luminosity increases both 
of these selections are expected to become less useful as the pile-up
of additional events can cause  jets to appear in these regions as
well as  degrading the measurements of the jets used for tagging/vetoing.

 An estimate of the effects was obtained using a full simulation study
of pileup in the ATLAS detector. Pileup from the appropriate number 
of minimum bias events in up to 25 beam crossings before and 3 beam
crossings after $t = 0$ was generated, taking into account 
the response of the different calorimeters. 
Jets were then found 
using a jet finder with cone size $\Delta R=0.4$ or $\Delta R =0.2$,
and assigned to ranges of rapidity
\begin{itemize}
\item Forward: $\eta>2.0$
\item Backward: $\eta<-2.0$
\item Central:  $\abs{\eta}<2.$
\end{itemize}

\begin{figure}
  \begin{minipage}[t]{0.48\textwidth} 
        \dofig{8cm}{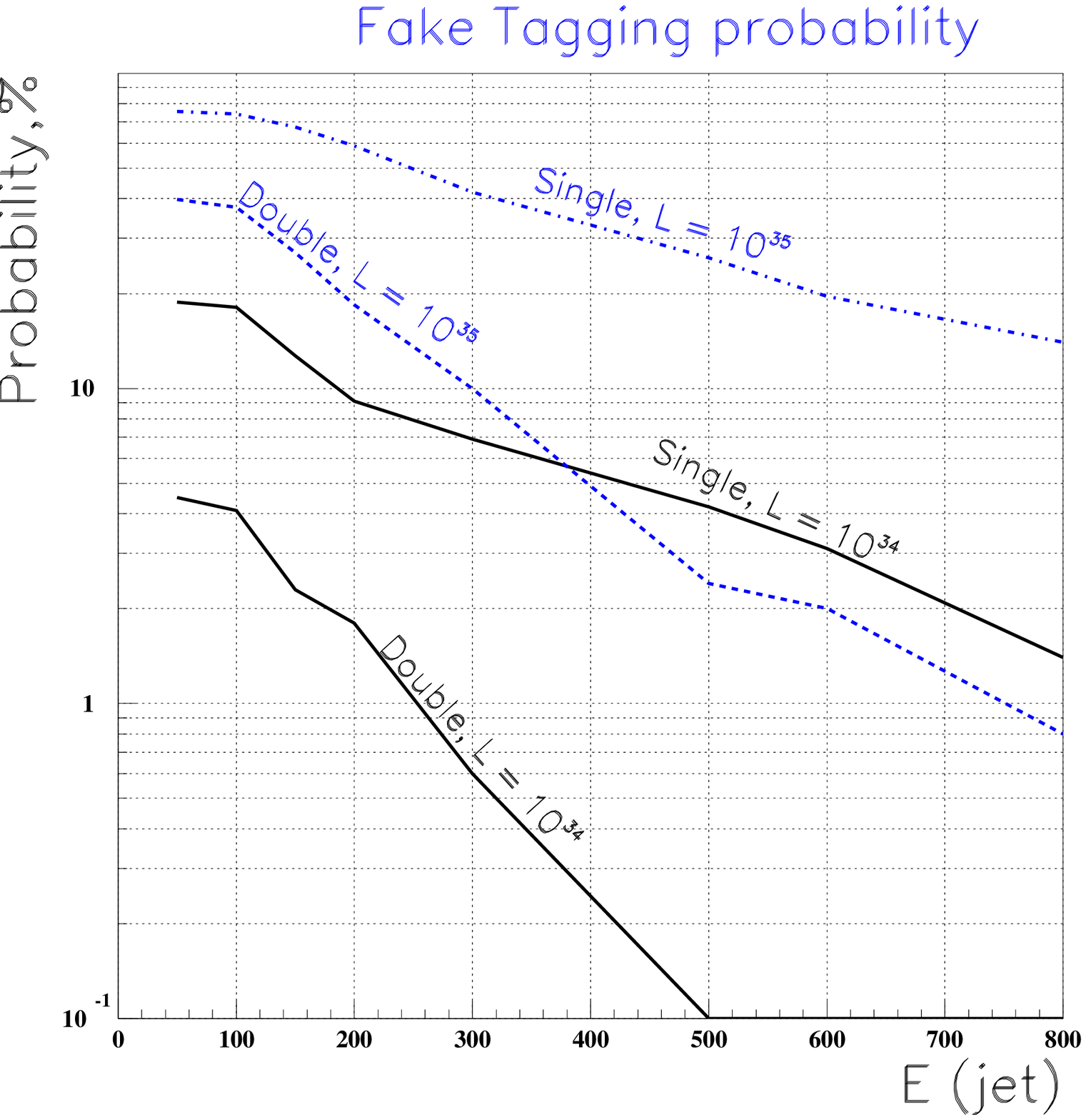}
  \end{minipage} \hfill
  \begin{minipage}[t]{0.48\textwidth} 
        \dofig{8cm}{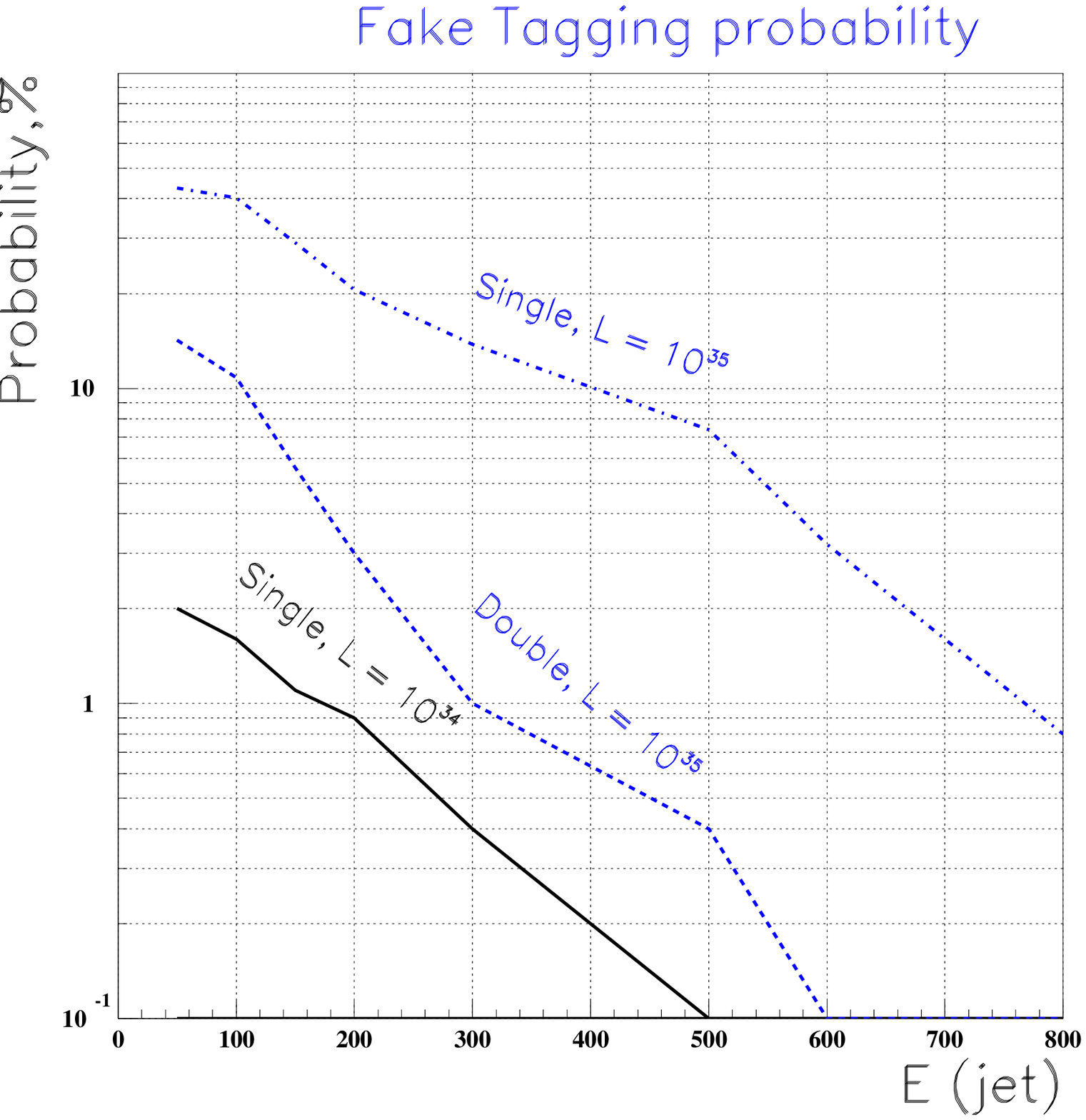}
  \end{minipage}
  \caption{
    Estimate of the probability of single and double forward jet tagging 
    from pile-up, for
    nominal and very high  luminosities, at 14 TeV, as a function of the
    jet energy, and for $\Delta R = 0.4$ (left) and $\Delta R = 0.2$ (right). \label{jjtag}}
\end{figure}

\begin{figure}
  \begin{minipage}[t]{0.48\textwidth} 
        \dofig{8cm}{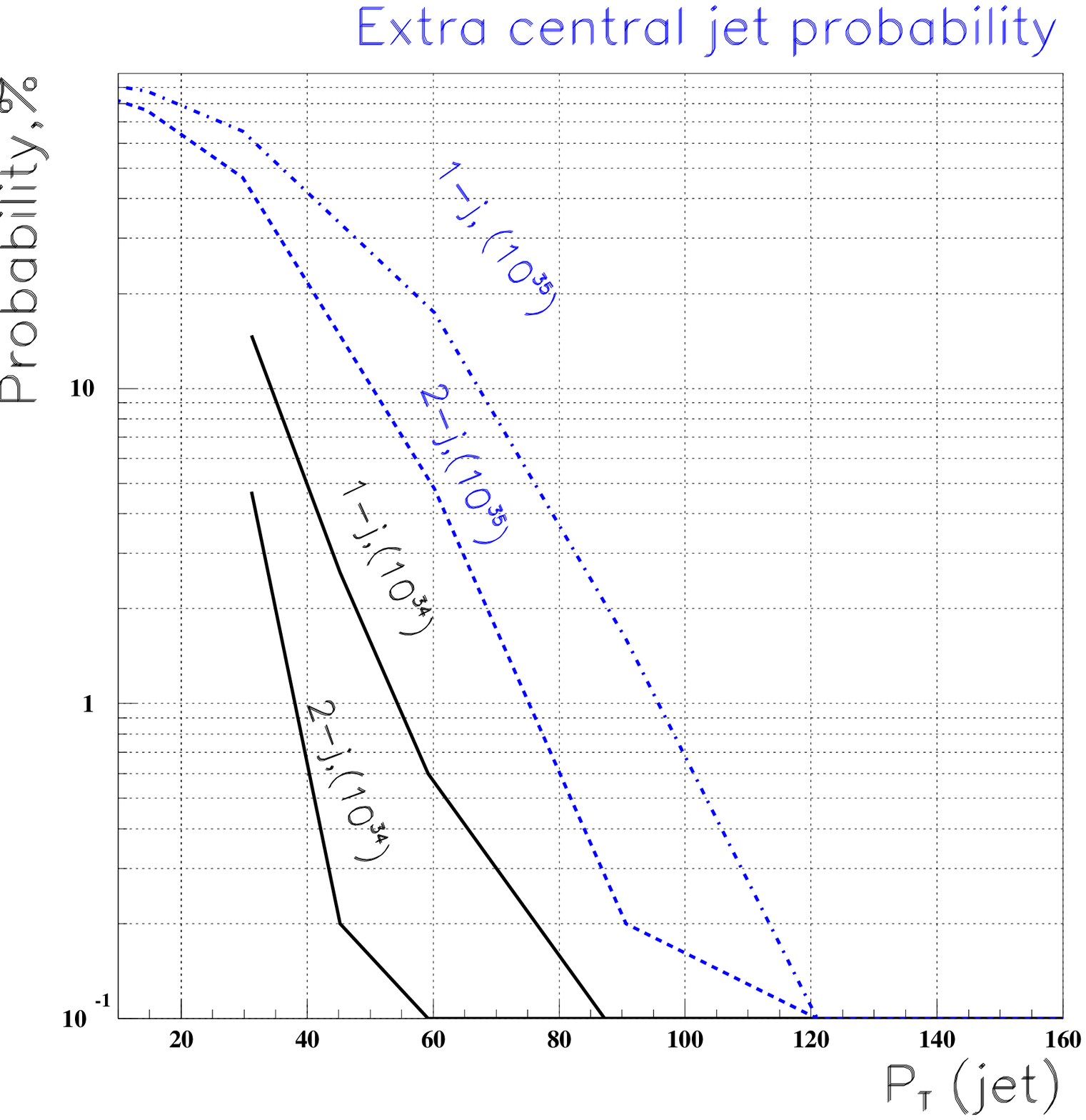}
  \end{minipage} \hfill
  \begin{minipage}[t]{0.48\textwidth} 
        \dofig{8cm}{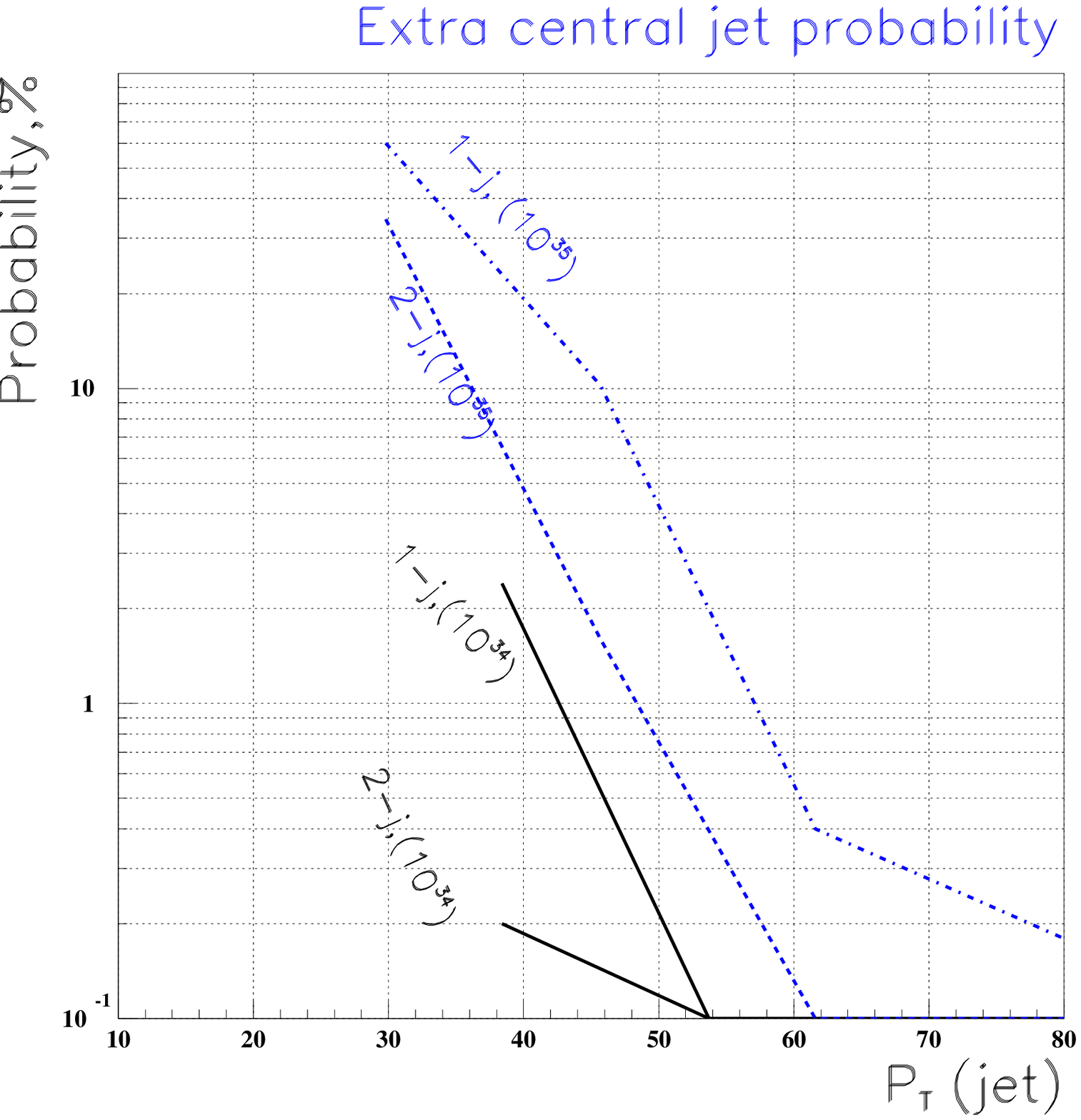}
  \end{minipage}
  \caption{
    Estimate of the probability of one or two extra central jets from pile-up, for
    nominal and very high  luminosities, at 14 TeV, as a function of the
    jet $p_T$ threshold, and for $\Delta R = 0.4$ (left) 
    and $\Delta R = 0.2$ (right).  \label{cjv} }
\end{figure}

A single tag is defined as an event with either a forward or backward
jet; a double tag has both.
The probability of an event consisting only of minimum bias
interactions having either a single or double 
jet tag is shown in Figure~\ref{jjtag} as a function of the jet energy.
The  probability of an event having an additional central jet is shown
in Figure~\ref{cjv} as a function of $p_T$. The values plotted in these 
figures are approximate, and depend sensitively on jet energy calibration
and reconstruction conditions.
At a luminosity
of $10^{35}$ cm$^{-2}$ sec$^{-1}$ the pile up of minimum bias events
renders forward jet tagging and central jet vetoing
difficult. The performance that is assumed here may be too pessimistic
as it may be possible to reduce the pile-up noise in the LAr
calorimeter by using optimal
filtering techniques at the upgraded luminosity
\cite{LATDR} and optimizing the jet algorithm.
At an energy of 28 TeV, the average number of minimum bias events increases
by about 15\% and the energies and multiplicities of particles
in these events is higher. In this case, as a rough approximation,
it will be assumed that the rates of
fake forward jet
tags and of extra central jets are twice as big as at nominal luminosity.

This tagging 
result has implications for strong $WW$ scattering signals as
is now discussed.

\section{Strongly coupled Vector Boson system}
If there is no light Higgs boson, then general
arguments\cite{arguments}
 imply that
scattering of electroweak gauge bosons at high energy will show
structure beyond that expected in the Standard Model. In order to
explore such signals it is necessary to measure final states of pairs
of gauge bosons with large invariant mass. 

\subsection{$W_L Z_L \to W_L Z_L$}

Estimates of the production of a $\rho$-like vector resonance of 
$W_LZ_L$ can be obtained
from the Chiral Lagrangian model, with the inverse amplitude method of
unitarization~\cite{Dobado}. The cross section depends, in
next to leading order, on a linear
combination $a_4-2a_5$ of two quadrilinear coupling parameters.
The model was implemented in PYTHIA. Only the channel
$W_L Z_L \to W_L Z_L \to \ell \nu \ell^+ \ell^-$ is
here considered, although the resonance can
be produced in the $q \overline{q}$ fusion channel at higher rate.
Forward jet tagging is here an essential ingredient to reduce the
background.

 The irreducible Standard Model background $ q q \to q q W Z$,
with transverse gauge bosons in the final state was generated with 
COMPHEP~\cite{COMPHEP} with cuts $p_T (q,W,Z) > 15$ GeV, and
$m_{WZ} > 500$ GeV, with CTEQ5L structure functions and $Q=M_Z$. 
The process includes electroweak and QCD
diagrams, as well as the quadrilinear gauge boson couplings. The
Higgs mass was set at the low value of 100 GeV, and the signal is
then defined, as in~\cite{Bagger95}, as the enhancement of the
SM prediction over the 100 GeV Higgs. Other background considered were
$Z b \bar b$ and $Z t \bar t$, also generated with COMPHEP, with
cuts $p_T(b,t)>15$ GeV and $p_T(Z) > 50$ GeV, and 
SM production of $WZ, ~ZZ$, generated with PYTHIA. Table~\ref{tab:WZback}
gives the cross sections for the different backgrounds.

\begin{table}
  \begin{center}
    \begin{tabular}{|c|c|c|c|c|}\hline
  Process         &  $\sigma$ (pb) \\ \hline
 $qq \to qqWZ$    &  1.45          \\
 $Z b \bar b$     &  141           \\
 $Z t \bar t$     &  2.23          \\
 $q q \to WZ$     &  3.00          \\
 $q q \to ZZ$     &  0.81          \\  \hline
    \end{tabular}
   \caption{ Cross sections for backgrounds to the $W_L Z_L \to W_L Z_L$ process
}
   \label{tab:WZback}
  \end{center}
\end{table}

 The selection criteria required leptonic cuts:
\begin{eqnarray}
 p_T(\ell_1) > 150 \mbox{GeV}, ~~~~p_T(\ell_2) > 100 \mbox{GeV}, ~~~~p_T(\ell_3) > 50 \mbox{GeV} \nonumber \\
 |m(\ell_1\ell_2) -m(Z)| < 10 \mbox{GeV} \nonumber \\
 \etmiss > 75 \mbox{GeV} \nonumber
\end{eqnarray}

and forward jet tagging, i.e. the presence of one forward and one backward jets ($|\eta| > 2$) with
energy greater than 300 (400) GeV at nominal (very high) luminosities, as well as a veto
if a jet with transverse momentum greater than 50 (70) GeV at nominal (very high) luminosity
was present in the central region ($|\eta| < 2$).

\begin{figure}
  \begin{minipage}[t]{0.48\textwidth} 
        \dofig{3in}{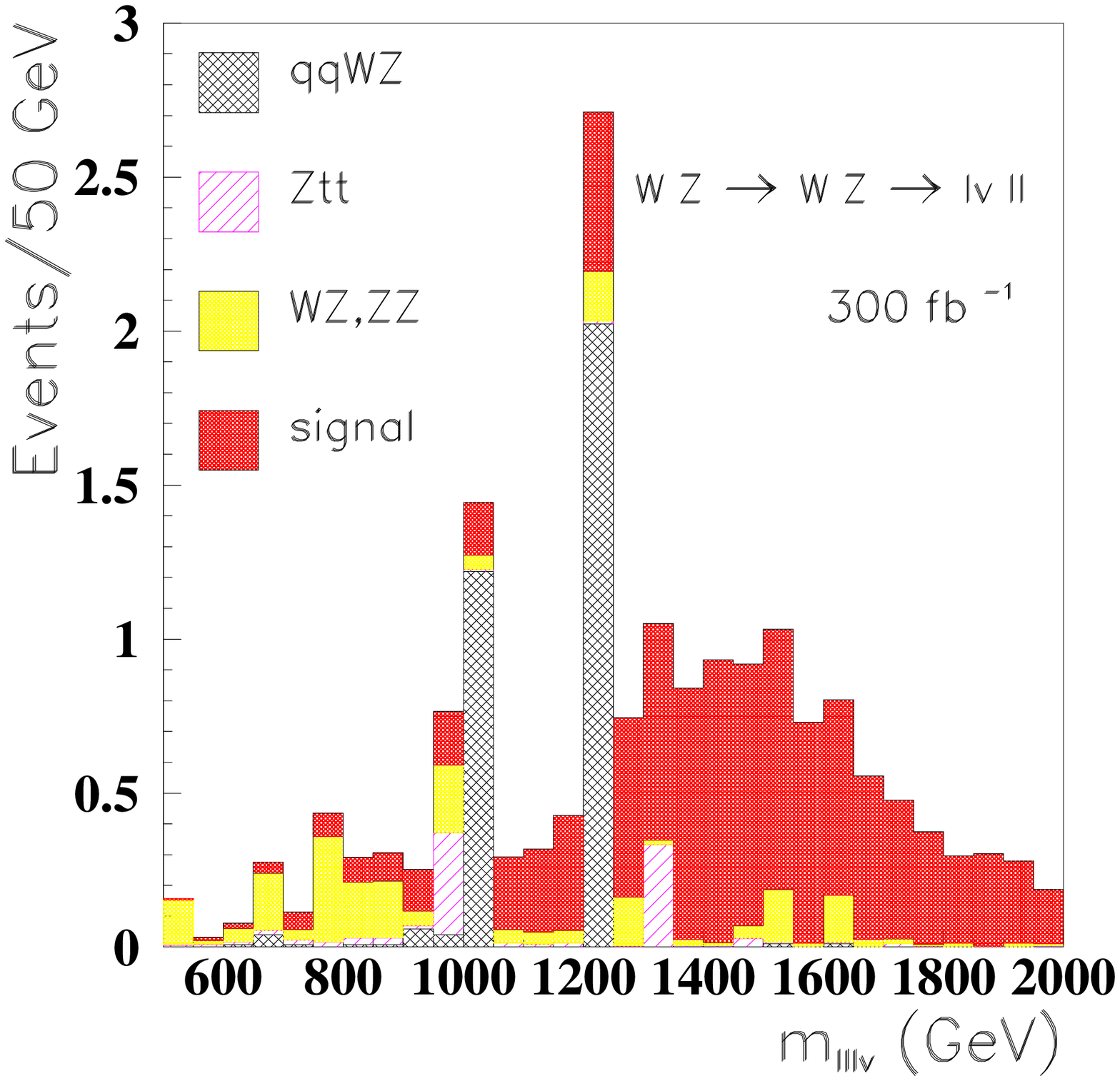}
  \end{minipage} \hfill
  \begin{minipage}[t]{0.48\textwidth} 
        \dofig{3in}{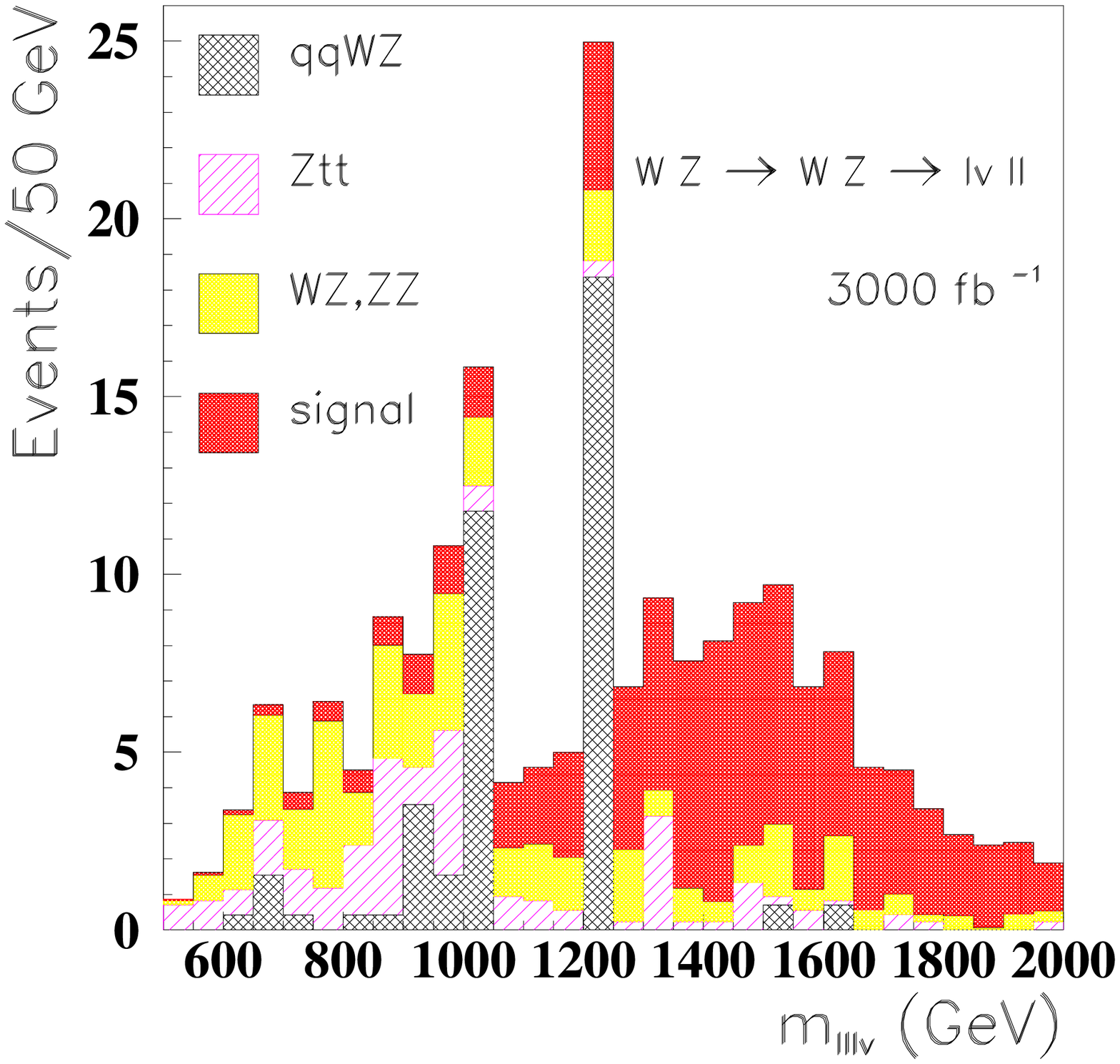}
  \end{minipage}
  \caption{
    Expected signal and background for a 1.5 TeV WZ resonance in the leptonic decay channel with
300 fb$^{-1}$ (left) and 3000 fb$^{-1}$ (right).
    \label{fig:WZ}}
\end{figure}

Fig.~\ref{fig:WZ} shows the expected signal for a 1.5 TeV resonance,
corresponding to a choice of Chiral Lagrangian parameters $a_4-2a_5=0.006$, 
at nominal LHC
and at an upgraded LHC with tenfold luminosity, taking into account the effects
of pileup according to Figs.~\ref{jjtag} and~\ref{cjv}, with $\Delta R = 0.4$.
The resonance is at the limit of
observation at LHC, with 6.6 events expected over a background of about 2.2
around the region of the peak, but at very high luminosity the signal 
has an approximate significance $S/\sqrt{B} = 48/\sqrt{16} = 12$.

\subsection{$Z_L Z_L$ scalar resonance}

Production of a scalar $Z_LZ_L$ resonance decaying by the golden
channel $ZZ \to 4 \ell$
is a rare process, well suited to an upgraded LHC. As for the $WZ$ 
resonance above, the Chiral Lagrangian
model, with regularization by the inverse amplitude method~\cite{Dobado}
was used.  The cross section depends on a
linear combination, $7a_4+11a_5$, of the same parameters as those
of the vector resonance and therefore observation of the scalar resonance
will help resolve  $a_4$ and $a_5$ unambiguously.
The $ZZ$ production signal occurs through the scattering processes
$W^+_LW^-_L \to Z_LZ_L$ and $Z_LZ_L \to Z_LZ_L$. Standard Model
background, leading to $qqZZ$ in the final state have been generated
with COMPHEP, with cuts $p_T (q,Z) > 15$ GeV, 
$m_{ZZ} > 500$ GeV, $m_{qq}>200$ GeV, with CTEQ5L as parton
distribution function and $Q=m_Z$. The process was
implemented in PYTHIA as an external process. 

The Higgs mass was set at 100 GeV so that the contribution from
longitudinal vector boson scattering was negligible in this background. 
With these conditions, the SM cross section is 69.4 fb. Other backgrounds
considered were $qq \to ZZ$, with cuts $m_{ZZ} > 500$ GeV,
and cross section 8.66 fb. The background $gg \to ZZ$ was not included,
but is expected to contribute about one third of the $q\bar q$ fusion
process~\cite{Dobado}.

\begin{figure}
\dofig{8cm}{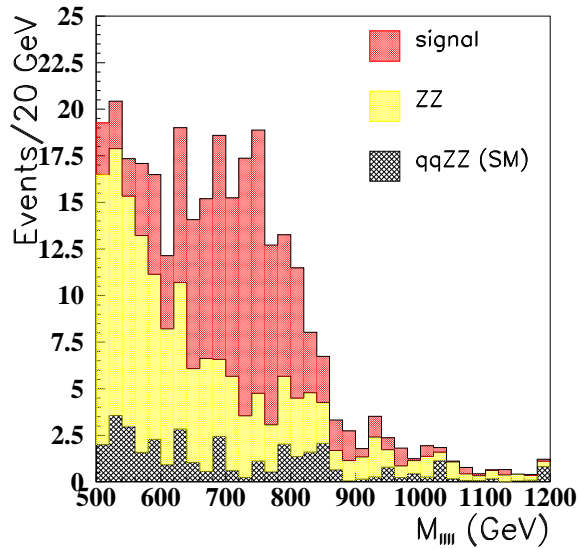}
\caption{Reconstructed ZZ mass from the leptonic decay of a scalar resonance
of 800 GeV. \label{fig:zz}}
\end{figure}

The analysis requires the presence of four isolated leptons having a transverse
momentum greater than 30 GeV, and leading to the reconstruction of two
$Z$'s. The identification efficiency of electrons was taken to be 80\%
and those of muons 90\%. Forward jet tagging was applied by requiring the presence of one
forward and one backward jet ($|\eta|>2$) with energies greater than 
400 GeV. No central jet veto was imposed, as the main backgrounds do
not justify it. Fake tagging by pileup was taken into account, according 
to the results in Fig.~\ref{jjtag}, with $\Delta R = 0.4$. 
With an integrated luminosity of 3000 fb, for
a 14 TeV collider, the expected signal and background for a resonance 
at 750 GeV, corresponding to $7a_4+11a_5 = 0.063$ is shown 
in Fig.~\ref{fig:zz}. Such a process would not be 
observable at the nominal LHC.

\subsection{$W^+_LW^+_L$}
Non-resonant production of vector boson pair at high mass presents a
challenge at LHC because the level of background needs to be very
well understood.

Production of $W^+W^+$ pairs has no contribution from $q\overline{q}$ fusion.
Two  models have been
considered:
\begin{itemize}
\item A Higgs boson of mass 1 TeV (as a reference point).
\item WW production with K-matrix unitarization.
\end{itemize}

Backgrounds arise from gauge boson pairs produced by  electroweak and gluon
exchange diagrams  \cite{TDR}. Here the  promising $\ell^+\ell^+\nu\nu$ final
state arising from $W^+W^+$ production is studied.
The analysis cuts are essentially the same
as in \cite{TDR}, consisting of leptonic cuts
\begin{itemize}
 \item two leptons ($e$ or $\mu$) reconstructed in the central region, $|\eta|<1.75$,
of positive charge and having $p_T > 40$ GeV
 \item The difference in azimuthal angle of the two leptons,
$cos(\Delta \phi (\ell\ell)) < -0.5 ~(-0.8)$ for  14 (28) TeV collisions
 \item $\Delta p_t(ll) =|\vec{p}_T(\ell_1)-\vec{p}_T(\ell_2)| > 100$ GeV for the 28 GeV case.
 \item missing transverse energy $E_T^{miss}>40 ~(50)$ GeV  for  14 (28) TeV collisions
\end{itemize}    

The jet tagging and vetoing cuts were varied as a function of the
energy and luminosity as shown in Table~\ref{tabcuts}.
Two forward tagging jets with energies above the amount
shown in the table were required and events were vetoed if there was a
central jet with transverse momentum greater than the indicated
amount. 
Table~\ref{wwstrong} shows the number of events expected 
and their significances for a typical
 set of cuts, for 
300 fb$^{-1}$ at nominal luminosity and 3000 fb$^{-1}$ at 
very high luminosity at 14 and 28 TeV.
 In the case of ultra high luminosity electron identification efficiency is
assumed to be 80\%.  It may be possible to reduce the
effects of pile up by improving filtering and optimising the jet
tagging.

\begin{table}
\begin{center}
\begin{tabular}{|c|c|c|c|c|}\hline
 & 300 fb$^{-1}$ & 3000 fb$^{-1}$
 & 300 fb$^{-1}$ & 3000 fb$^{-1}$\cr

 &14 TeV&14 TeV& 28 TeV&
28 TeV\cr\hline
Forward tag $E$&
400 &
600&
600&
800\cr
Forward tag $p_T$&
90 &
90 &
150&
150\cr
Central veto&
40&
60&
50&
100
\cr\hline
\end{tabular}
\caption{Energies of the two forward jets required for tagging and the
  transverse momentum of the central jets used for vetoing in GeV for
  the various energies and luminosities in the study of $W^+W^+$ final
  states. The forward jets were also required to have a transverse 
  momentum less than shown.
}\label{tabcuts}
\end{center}
\end{table}

\begin{table}
\begin{center}
\begin{tabular}{|c|c|c|c|c|}\hline
 & 300 fb$^{-1}$ & 3000 fb$^{-1}$
 & 300 fb$^{-1}$ & 3000 fb$^{-1}$\cr

Model &14 TeV&14 TeV& 28 TeV&
28 TeV\cr\hline
Background&
7.9&
44&
20&
180\cr
K-matrix Unitarization&
14&
87&
57&
490\cr
Significance&
3.0&
7.6&
6.5&
18.9\cr
Higgs, 1 TeV &
7.2&
42&
18&
147\cr
Significance&
1.8&
4.5&
2.9&
8.1\cr\hline
\end{tabular}
\caption{Numbers of reconstructed events above an invariant
mass of 600 GeV (for 14 TeV) and 800 GeV (28 TeV) for models of a strongly
  coupled Higgs sector and for the background.
The significance was computed as $S/\sqrt{S+B}$.}\label{wwstrong}
\end{center}
\end{table}

\begin{figure}
  \begin{minipage}[t]{0.48\textwidth} 
        \dofig{8cm}{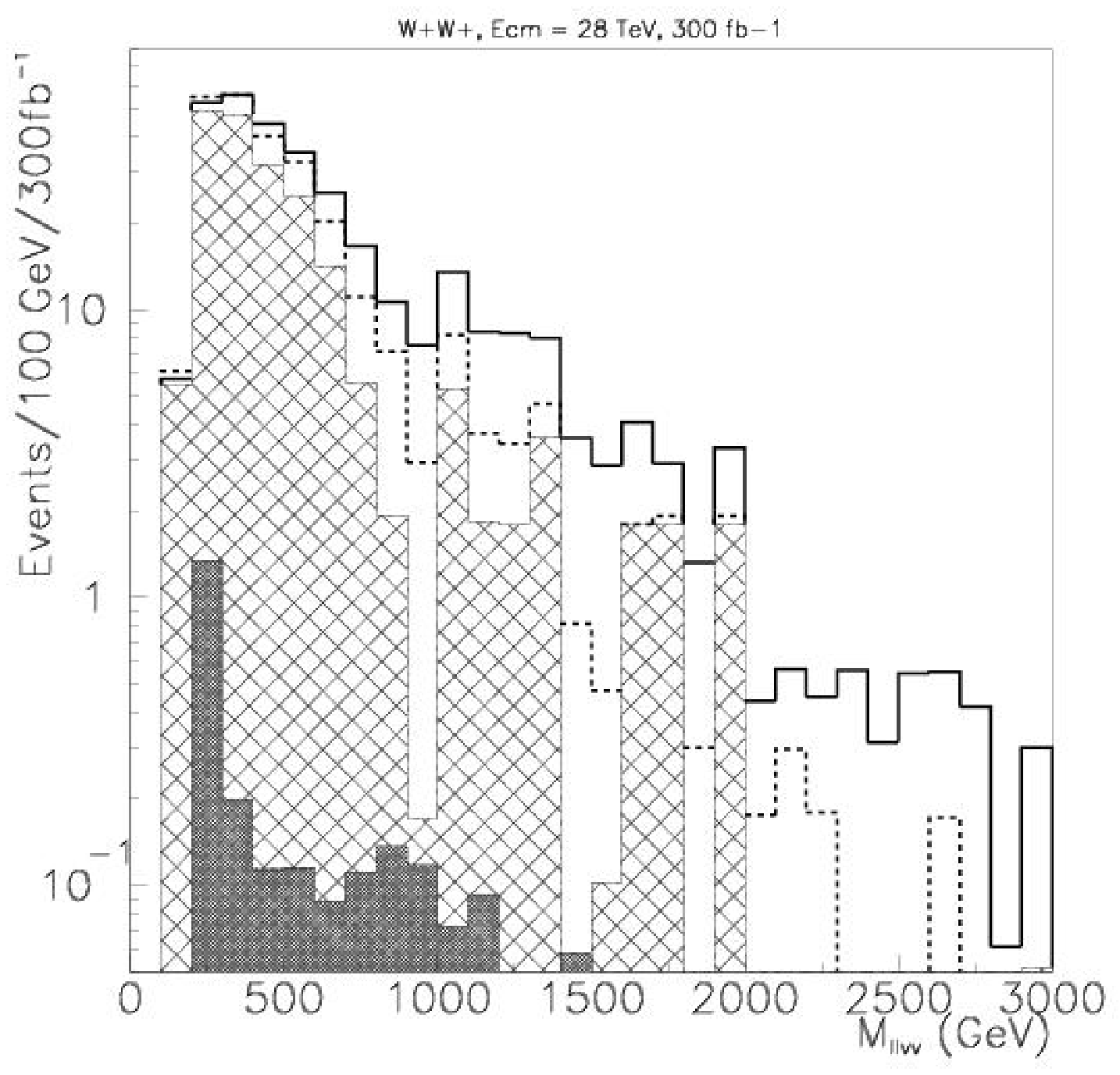}
  \end{minipage} \hfill
  \begin{minipage}[t]{0.48\textwidth} 
        \dofig{8cm}{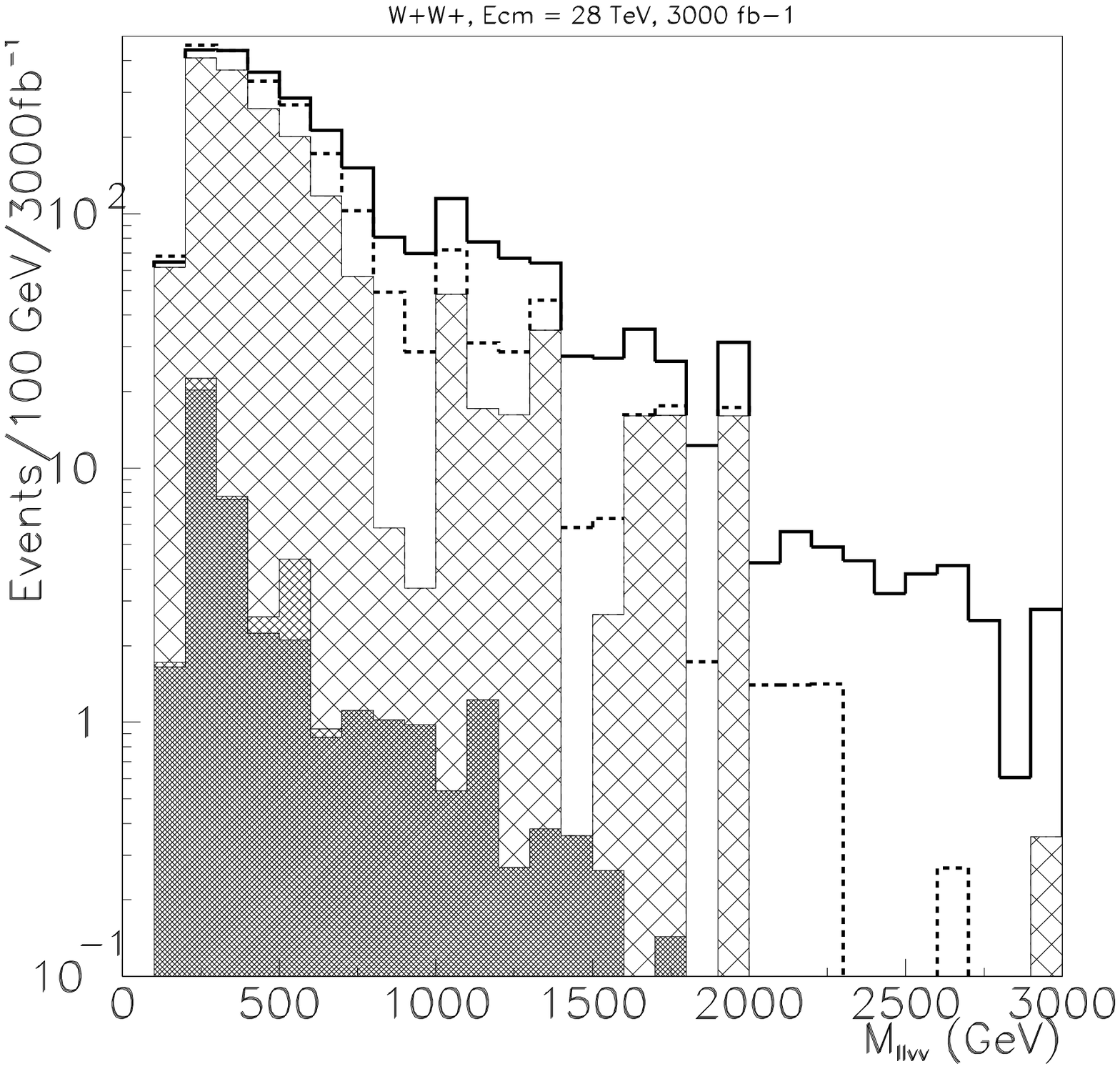}
  \end{minipage}
  \caption{
Signals from strong WW scattering at 28 TeV and 300 fb$^{-1}$ (left)
and 3000 fb$^{-1}$ (right) of
  integrated luminosity. Event rates are
   shown as a function of the invariant mass of the
   $\ell\ell \etmiss$ system.
 The backgrounds are shown as histograms, 
 from inside to outside (or darker to lighter): 
   WZ continuum:
   W pairs (transverse) from gluon exchange diagrams:
   W pairs (transverse) from  electroweak ($\gamma$ and Z exchange) diagrams. 
 The possible signals are:  continuous line,  K-matrix unitarization:
   dashed line: Higgs, 1 TeV.
 \label{wwfigh}}
\end{figure}

The increase in energy to 28 TeV would enhance considerably the
sensitivity to high 
mass vector boson pair production, even in a non-resonant scenario. 
The study of the signal  could be pushed from a limit of about 1
TeV in the invariant mass of the $WW$ system to 
around 1.5 TeV.
 Fig.~\ref{wwfigh} shows the 
signal and background
 distributions as a function of the invariant mass formed from the two 
 leptons and the missing transverse momentum ({\it c.f. }Figure 19-110
   of \cite{TDR}). In spite of the apparent increase 
in statistical significance, the systematic errors are likely to be
 very high since the shapes of the backgrounds and signal are similar. 
The extraction of  signals should be done 
 with many different choices of cuts so as
to verify the dependence  on the cuts and 
constrain the systematics.  This analysis
can only be an estimate of the reach. At high 
luminosities  a good understanding of jet
reconstruction and backgrounds will be needed.

%

\section{Extra dimensions}

One of the possible signals expected from large extra dimensions is the production of
 jets or
photons in association with missing transverse energy. The former
channel is more sensitive so will be considered here. The background
is dominated by the final state $Z(\to \nu\nu)+jets$ ~\cite{vaca}. 

In order to assess the impact
of the possible upgrades a  study using ATLFAST\cite{atlfast} was made. The signal is
characterized by two parameters, the number of extra dimensions
$\delta$ and the scale $M_D$ characterizing the scale of gravity. 
Table~\ref{extradtab} shows the maximum value of  $M_D$ that can be
detected for a given value of $\delta$ in the four scenarios.

\begin{table}[htbp]
\begin{center}  \begin{tabular}{|c|c|c|c|c|}\hline
$\delta$&14 TeV& 14 TeV &28 TeV& 28 TeV\cr
 & 100 fb$^{-1}$ & 1000 fb$^{-1}$
 & 100 fb$^{-1}$ & 1000 fb$^{-1}$\cr
\hline
2&9 & 12& 15&19\cr\hline
3&6.8 & 8.3& 11.5 &14\cr\hline
4&5.8 & 6.9& 10 &12\cr\hline
  \end{tabular}
\end{center}
\caption{ 5 $\sigma$ discovery reach  on $M_D$, in TeV,  as a function
  of the number of extra dimensions ($\delta$) for various values of energy and integrated luminosity.}\label{extradtab}  
\end{table}

It can be seen from the table that doubling the LHC energy to 28 TeV
approximately doubles the reach in $M_D$ for any value of $\delta$. An
increase in the integrated luminosity by a factor of 10  raises the
reach by approximately 30\%. If a signal is observed at 14 TeV, the
energy increase is such that model dependent signals can be expected
to be observed at 28 TeV. Such observations would therefore provide valuable
insight into the dynamics of the underlying theory. Detector
performance is not expected to be critical as 
events with jets and missing energy in the TeV range are relevant for this search.

\section{Searches for New Gauge Bosons}

\begin{figure}
\dofig{3in}{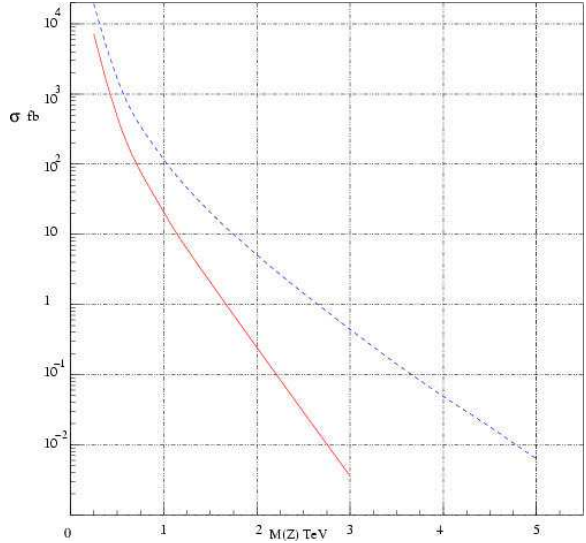}
\caption{Production rate for  $Z^{\prime}\to \mu^+\mu^-$ at 14 and 28 TeV as a
  function of the $Z^{\prime}$ mass.
 \label{zprime}}
\end{figure}

As an example of the impact of the possible upgrades, 
Figure \ref{zprime} shows the production rate for $\mu^+\mu^-$ final states
arising from the decay of a $Z^{\prime}$ gauge boson with the same
couplings to quarks and leptons as the Standard Model $Z$ as a function of the mass of the
$Z^{\prime}$. The final state $e^+e^-$ is not considered, because it
 is more difficult to exploit at the highest luminosity
particularly if the tracker is not upgraded. The calorimeter can still be used to detect this channel but the signs of the electrons cannot be determined and measurements of asymmetries cannot be undertaken.  If this
channel were included and had the same acceptance as the muon
channel, the mass reach would increase by approximately 20\%.

The discovery limit for $Z^{\prime}\to \mu^+\mu^-$  is given in
Table~\ref{zprime-tab} from which it can be seen that the combination
of energy and luminosity upgrade doubles the mass reach.

\begin{table}[htbp]
\begin{center}  
\begin{tabular}{|c|c|c|c|}\hline
14 TeV& 14 TeV &28 TeV& 28 TeV
\cr  100 fb$^{-1}$ & 1000 fb$^{-1}$&
 100 fb$^{-1}$ & 1000 fb$^{-1}$\cr\hline
4.5 & 5.4& 7.0 &9.5\cr\hline
  \end{tabular}
\end{center}
\caption{$5\sigma$ discovery reach in   $Z^{\prime}$ mass in TeV in the
  $\mu^+\mu^-$ final state, for a
  $Z^{\prime}$ with the same couplings to quarks and leptons as the
  Standard Model $Z$. }\label{zprime-tab}  
\end{table}

The decay  $Z^{\prime} \to WW\to \ell\nu jj$  
(see Section 21.6.1.4 of \cite{TDR})
 has also been
studied. This channel is not competitive for discovery but can provide
important information on the couplings of  the $Z^{\prime}$. This study
uses the final state where the lepton is a muon, as electron detection
will be more difficult at higher luminosity. The mass
reachable in this final state is shown in Table~\ref{zprime-tab2}.

\begin{table}[htbp]
\begin{center}
  \begin{tabular}{|c|c|c|c|c|}\hline
      & 14 TeV 300 fb$^{-1}$&  14 TeV  3000 fb$^{-1}$
&28 TeV 300 fb$^{-1}$ &28 TeV 3000 fb$^{-1}$
\cr\hline
Mass (TeV)&1.9 & 2.5& 2.4& 3.3\cr\hline
  \end{tabular}\end{center}
\caption{ $5\sigma$ discovery reach in TeV for  a $Z^{\prime}$   observed in
  the $WW\to \mu\nu j j$ final state.}\label{zprime-tab2}  
\end{table}

The ability to measure muons and jets of high transverse momenta is
needed for these channels which is therefore 
not very sensitive to
performance degradations at  higher luminosity. 

\section{Searches for excited quarks}

\begin{figure}
\dofig{4in}{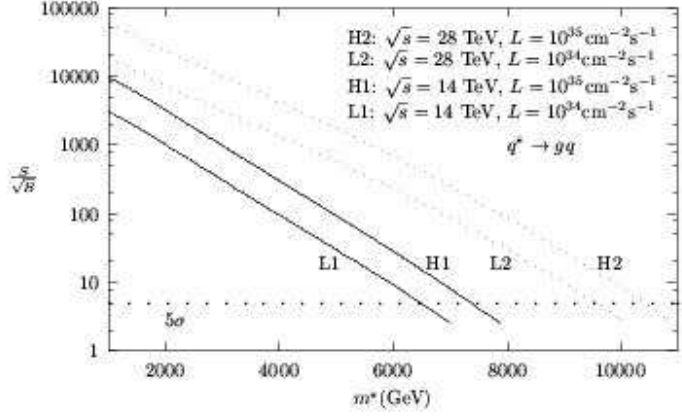}
\caption{The statistical significance of an excited
  quark signal as a function of the excited quark mass in the final
  state $jet+jet$ at 14 and 28 TeV  and integrated luminosities of 100 and 1000 fb$^{-1}$.
 \label{excite1}}
\end{figure}

\begin{figure}
\dofig{4in}{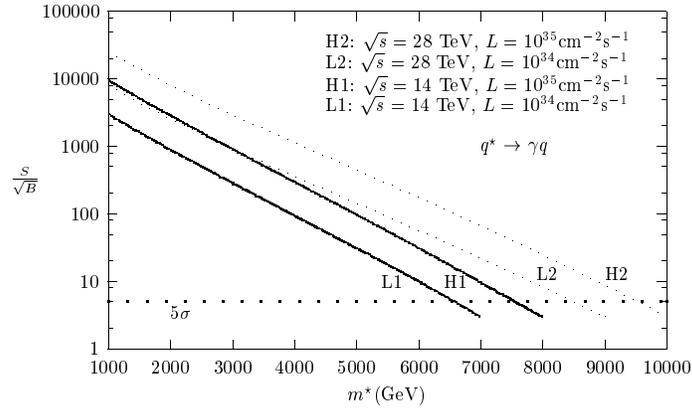}
\caption{The statistical significance of an excited
  quark signal as a function of the excited quark mass in the final
  state $\gamma+jet$ at  14 and 28 TeV and integrated luminosities of 100 and 1000 fb$^{-1}$.
 \label{excite2}}
\end{figure}

Section 21.3 of Ref~\cite{TDR} and Ref.~\cite{Cakir} discuss signals for new exotic
quarks. Some of these studies have been repeated for the upgrade
scenarios.
 The
decays $q^*\to qg$ and $q^*\to q \gamma$ have been
simulated. Figures~\ref{excite1} and \ref{excite2} show the
statistical significance of the signal in these channels as a function 
of the excited quark mass in the $jet-jet$ and $jet-\gamma$ final
states.
The mass reach increases by about 4 TeV for the highest
energy/luminosity configuration. As only jets and photons of the very
highest transverse energy are needed in the analysis, this  process is
rather robust even at the very highest luminosity.

\section{Triple gauge-boson couplings}

\begin{figure}
\dofig{6in}{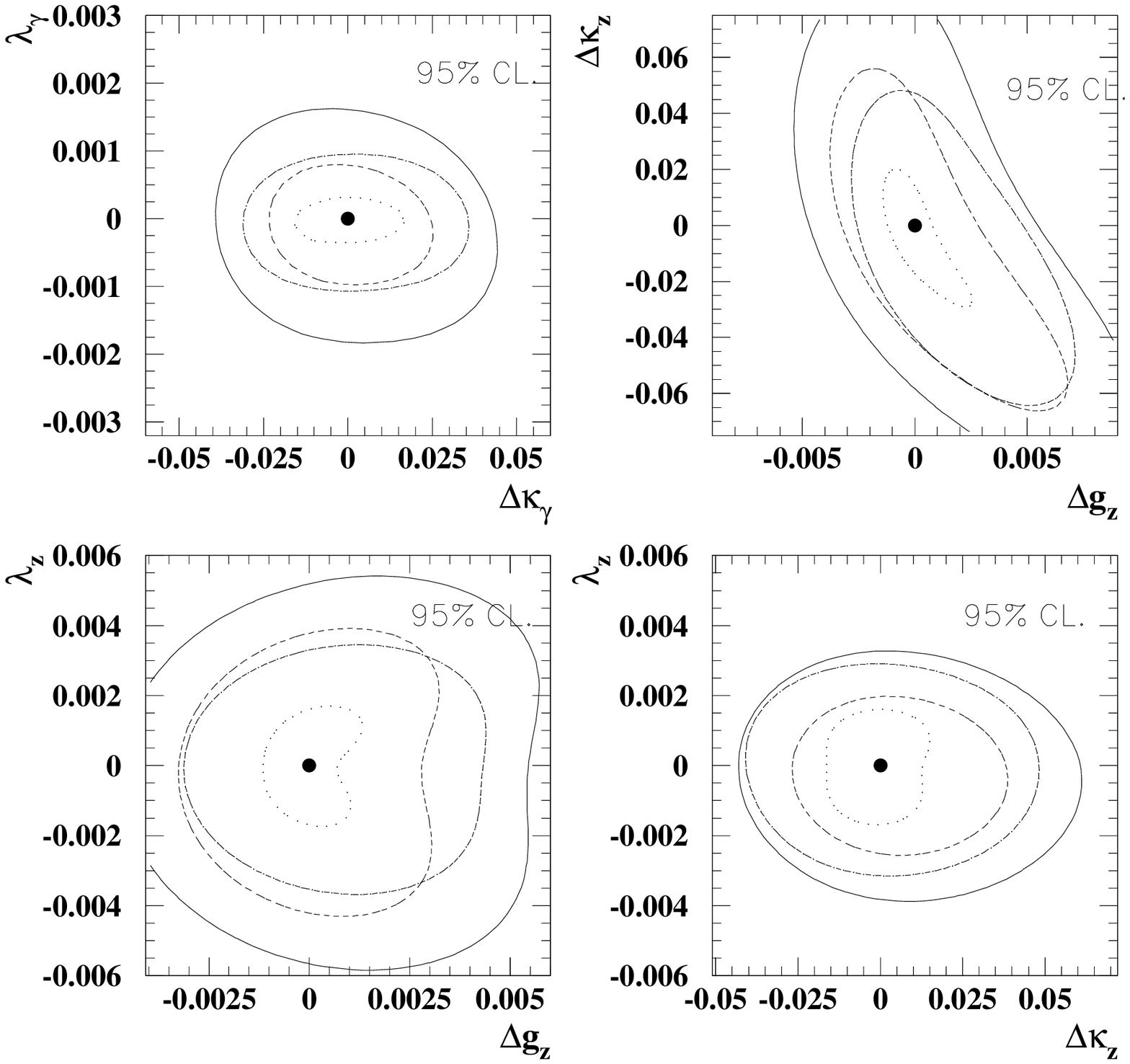}
\caption{Expected 95\% C.L. constraints on triple gauge-boson couplings 
 $\kappa$
  and $\lambda$ resulting from two parameter fits. The contours  correspond
  to 14 TeV and  100 fb$^{-1}$ (solid), 28 TeV and 100 fb$^{-1}$ (dot
  dash), 14 TeV and 
  1000 fb$^{-1}$ (dash), 28 TeV and 1000  fb$^{-1}$ (dotted).
 \label{triple}}
\end{figure}

The final state $W\gamma\to\ell\nu\gamma$ ($WZ\to \ell\nu\ell\ell$)
has been used to probe the
couplings $\lambda_{\gamma}$ and $\Delta\kappa_{\gamma}$ (
$\lambda_{Z}$,
 $\Delta\kappa_{Z}$ and $g_1^Z$)
that describe 
the $WW\gamma$ ($WWZ$) interaction vertex. More details of the methodology can 
be found in Section 16.2  of \cite{TDR}. For a luminosity of $10^{35}$
cm$^{-2}$ sec$^{-1}$, the analysis reported here uses only muons and
photons. This represents a loss of 50\% (75\%) of the $W\gamma$ ($WZ$) 
effective rate. The constraints on the $\kappa$ couplings arise
primarily from the angular distributions whereas those on the
$\lambda$ couplings arise from the transverse momentum
distributions. As only the latter are used here, the results are
pessimistic in the case of the  $\kappa$ couplings.
The expected sensitivity is shown in Table~\ref{triple-tab} and the
correlations in Fig.~\ref{triple}. Note that 
both  the energy or luminosity upgrades extend the sensitivity for
$\lambda_{\gamma}$ into
the range ($\sim$ 0.001) expected from radiative corrections 
in the Standard Model and should allow a
meaningful test of these corrections and others that arise for example
in supersymmetric models. 

In contrast to most of the other examples discussed, the
luminosity increase is more powerful than the energy increase in this case.    

\begin{table}
\begin{center}\begin{tabular}{|c|c|c|c|c|}
\hline
Coupling&14 TeV& 14 TeV &28 TeV& 28 TeV\cr\hline
&100 fb$^{-1}$&1000 fb$^{-1}$&100 fb$^{-1}$&1000 fb$^{-1}$\cr\hline
$\lambda_\gamma$&0.0014&0.0006&0.0008&0.0002\cr\hline
$\lambda_Z$&0.0028&0.0018&0.0023&0.009\cr\hline
$\Delta \kappa_{\gamma}$&0.034&0.020&0.027&0.013\cr\hline
$\Delta \kappa_Z$&0.040&0.034&0.036&0.013\cr\hline
$g_1^Z$&0.0038&0.0024&0.0023&0.0007\cr\hline
\end{tabular}
\end{center}
\caption{ 59\% C.L. constraints on the triple gauge boson couplings}
\label{triple-tab}
\end{table}

\section{Comments and conclusion}

Although there are other processes where some impact from the upgrades can be
expected and which have not been studied, this survey of physics processes at an upgraded LHC enables one to
draw certain conclusions.


Without major detector upgrades, {\it i.e.} by using only  final states of 
high $p_T$ jets, photons and muons,  a luminosity upgrade is expected to 
provide a $\sim 20$\% improvement in the mass reach for new  physics as SUSY 
and extra-dimensions. This increase is significant for signals at the
limit 
of the LHC sensitivity. However,
the failure of the Inner Detector will
compromise electron identification as well as b and hadronic $\tau$
tagging. For many new physics processes, statistical samples are
improved  by combining channels with electrons and muons. In addition
valuable information about new physics can be obtained by looking for
violation of $e/\mu/\tau$ universality.  Therefore major detector upgrades would be needed to exploit fully the factor of ten increase in luminosity.
Pile-up of minimum bias events has severe impact on the
ability to use forward jet tagging and central jet vetoing as a tool
to enhance signal to background ratios at luminosities significantly 
higher than $10^{34}$ cm$^{-2}$ sec$^{-1}$.

 The energy upgrade would be  much easier to exploit. It could significantly
 enhance the physics reach of the LHC by almost  a factor of two in
 mass.
 In addition, if new physics is discovered, then the energy 
upgrade would allow significant further
 study of the new physics. However, this scenario remains speculative.

\end{document}